\documentclass{article}
\pdfoutput=1

\usepackage{amsmath}
\usepackage{amssymb}
\usepackage{graphicx}
\usepackage{dsfont}

\newcommand{\rem}[1]{}

\newcommand{\todo}[1]{}

\newcommand{\dof}{DoF}

\newcommand{\sphere}{\mathbb{S}}

\newcommand{\nhim}{{\sphere_{\text{NHIM}}^{2n-3}(h)}}
\newcommand{\ts}{{\sphere_{\text{ds}}^{2n-2}(h)}}

\newcommand{\tsf}{{B_{\text{ds,\,f}}^{2n-2}(h)}}
\newcommand{\tsb}{{B_{\text{ds,\,b}}^{2n-2}(h)}}
\newcommand{\po}{{\sphere_{\text{NHIM}}^1(h)}}

\newcommand{\tstwoDOF}{{\sphere_{\text{ds}}^{2}(h)}}
\newcommand{\tsftwoDOF}{{B_{\text{ds,\,f}}^{2}(h)}}
\newcommand{\tsbtwoDOF}{{B_{\text{ds,\,b}}^{2}(h)}}

\newcommand{\capsty}{\footnotesize}

\newcommand{\Capts}[1]{}
\newcommand{\FIGo}[3]{\begin{figure}%
#3% % comment this line out to remove all figures
\caption[]{\capsty #2}%
\label{#1}%
\end{figure}}

\newcommand{\ui}{\mathrm{i}}
\newcommand{\ue}{\mathrm{e}}

\newcommand{\T}{\mathds{T}}
\newcommand{\R}{\mathds{R}}

\newcommand{\M}{\mathcal{M}}

\newcommand{\ccp}{{\mathcal P}}

\providecommand{\norm}[1]{\lVert#1\rVert}

\begin{document}

\bibliographystyle{natbib}

\title{Geometrical Models of the Phase Space Structures Governing Reaction Dynamics}

\author{
        Holger Waalkens$^{1,2}$
and Stephen Wiggins$^1$
   }

\maketitle

\noindent
{\small $^1$ School of Mathematics, University of Bristol, University Walk, Bristol BS8 1TW, UK}\\
{\small $^2$ Department of Mathematics, University of Groningen, Nijenborgh 9, 9747 AG Groningen, The Netherlands}\\[2ex]
E-mail: h.waalkens@math.rug.nl, s.wiggins@bristol.ac.uk

\begin{abstract}
Hamiltonian dynamical systems possessing equilibria of
$\mbox{saddle} \times \mbox{centre} \times\cdots\times
\mbox{centre}$ stability type display \emph{reaction-type
dynamics} for energies close to the energy of such equilibria; entrance and exit
from certain regions of the phase space is only possible via
narrow \emph{bottlenecks} created by the influence of the
equilibrium points.
In this paper we provide a thorough pedagogical description of the phase space
structures that are responsible for controlling transport in these
problems.
Of central importance is the existence of a \emph{Normally Hyperbolic
Invariant Manifold (NHIM)}, whose \emph{stable and unstable manifolds}
have sufficient dimensionality to act as separatrices, partitioning
energy surfaces into regions of qualitatively distinct behaviour.
This NHIM forms the natural (dynamical) equator of a (spherical)
\emph{dividing surface} which locally divides an energy surface into
two components (`reactants' and `products'), one on either side of the bottleneck.  
This dividing surface has all the desired properties sought for in 
\emph{transition state theory} 
where reaction rates are computed from the flux through a dividing surface.
In fact, the dividing surface that we construct is crossed exactly once by reactive trajectories, and not crossed by nonreactive trajectories, and related to these properties,  minimizes the flux upon variation of the dividing surface. 

We discuss three presentations of the energy surface and the phase space structures contained in it 
for 2-degree-of-freedom (DoF) systems in the threedimensional space $\R^3$, and two
schematic models  which capture 
many of the essential features of the dynamics for $n$-DoF systems.
In addition, we elucidate the structure of the NHIM. 

\end{abstract}

\tableofcontents

%-----------------------------------------------------------------------
\section{Introduction}
\label{sec:introduction}
%-----------------------------------------------------------------------

Transition state theory (TST)  is one of  the most important
theoretical and computational approaches to analyzing chemical
reaction dynamics, both from a qualitative and quantitative point
of view. Transition state theory was created in the 1930's, with most of
the credit being given to Eyring, Polanyi, and Wigner, who are
referred to as the ``founding trinity of TST'' in Miller's
important review on chemical reaction rates (\cite{Miller1}).
Nevertheless, important contributions were also made by Evans,
Farkas, Szilard,  Horiuti, Pelzer, and Marcelin, and these are
described in the discussions of the historical development of the
subject given in \cite{LaidlerKing, PollakTalkner}.

The central ideas of TST are fundamental and provide a
very natural and fruitful approach to ``transformation''  in many
areas far removed from chemistry. For example, it has been used
for certain studies in atomic physics \cite{JaffeFarellyUzer2000},
studies of the rearrangements of clusters \cite{Komats4}, studies
of transport phenomena in solid state and semi-conductor physics
\cite{Jacucci,Eckhardt95}, studies of diffusion dynamics in
materials \cite{vmg}, cosmology \cite{Olive}, and celestial
mechanics \cite{Jaffe,WaalkensBurbanksWiggins05b}.

The fundamental assumptions of TST we  clearly stated by 
Wigner, who begins by stating that he considers
chemical reactions in a setting where  the equilibrium
Maxwell-Boltzmann velocity and energy distributions are maintained
(see \cite{Mahan} for a detailed discussion of this point) and for
which the potential energy surface is known (\cite{Garrett}). He
then gives the following assumptions from which he derives TST:

\begin{enumerate}

\item the motion of the nuclei occurs on the Born-Oppenheimer
potential energy surface (``electronic adiabaticity'' of the
reaction)

\item classical mechanics adequately describes the  motion of the
nuclei

\item there exists a hypersurface in phase space dividing the
energy surface into a region of reactants and a region of products
having the property that all trajectories that pass from reactants
to products must cross this dividing surface precisely once.

\end{enumerate}

\noindent It is important to note that Wigner clearly developed
his ideas in {\em phase space}, the arena for dynamics. It is
important to keep this in mind since a great deal of later
developments occur in {\em configuration space}, in which certain
dynamical properties are obscured. Nevertheless, the typical mental picture for the geometry 
associated with TST  is built around
a saddle point of a potential energy surface for a one degree-of-freedom (DoF) or two
DoF system, i.e. it starts with a configuration space notion of
sufficiently low dimension that standard visualisation intuition
applies. 

It follows from Wigner's assumptions that fundamental geometric structure of TST is
a dividing surface which locally divides the energy surface into
two disjoint components (which we refer to as the ``bottleneck
property'')  and which is free of local ``recrossing'' (otherwise,
the flux through the dividing surface would be overestimated).
These properties are crucial for reaction rate computations.

The problem of how to define and construct a dividing surface with
these properties was solved for two degrees of freedom (\dof) in the 70's and 80's
by Pechukas, Pollak and others \cite{cp,Pechukas81,Pechukas1,Pechukas2,PechukasPollak77,
PechukasPollak79,Pechukas3,Pollak81,PollakPechukas79,PollakChild80} who
constructed the dividing surface from a periodic orbit, the so
called {\em periodic dividing surface} (PODS).

The generalization to systems with more than two degrees of
freedom has posed severe problems. Periodic orbits lack sufficient
dimensionality for the construction of dividing surfaces for
systems with three or more degrees of freedom.  Progress in constructing a dividing surface for systems with three or more degrees of freedom was made with the introduction of a fundamentally new object, a normally hyperbolic invariant manifold, or ``NHIM''
(\cite{Wiggins90, Wiggins94}), that takes the place of the periodic orbit. The
NHIM is the main building block (and the natural higher
dimensional generalization of the periodic orbit used by Pechukas,
Pollak and others) of the phase space transition state theory
mentioned above. This has been described more fully in 
recent work has shown that the fundamental
assumptions of transition state theory can only be realized in
phase space for systems with three or more DoF. This work is described in a series of papers 
\cite{wwju},\cite{ujpyw},
\cite{WaalkensBurbanksWiggins04}, \cite{WaalkensWiggins04},
\cite{WaalkensBurbanksWigginsb04},\cite{WaalkensBurbanksWiggins05},
\cite{WaalkensBurbanksWiggins05c},
\cite{SchubertWaalkensWiggins06}, \cite{WaalkensSchubertWiggins08}
where the fundamental dynamical and geometrical framework for phase
space TST is developed.
Morever, it is shown that new types
of ``phase space structures'' are responsible for governing the
dynamics.

Chief among these structures is the NHIM.
\footnote{The concept of an ``invariant manifold'' is fundamental to the description and quantification of a wide variety of dynamical phenomena. However, the use of the term, and its significance, may not be so clear to someone with a non-dynamical systems oriented background. Therefore we give a brief explanation of the term, from which will follow its implications for trajectories in phase space. For our purposes we can think of the term ``manifold'' as being synonomous with the term ``surface''. ``Invariance'' of the manifold can be characterized in two (equivalent) ways. One is that trajectories through initial conditions on the manifold must be contained in the manifold for their entire pasts and futures. The other way of characterizing an invariant manifold is by saying that the vector field giving rise to the dynamics is tangent to the manifold at every point of the manifold. It follows immediately from both characterizations of invariance that any trajectory with initial condition {\em not} on the invariant manifold can never ``cross''  the invariant manifold. 
In this way, invariant manifolds of codimension 1, i.e. of dimension one less than the dimension of phase space, act as separatrices. They form the  ``barriers to transport'' or ``impenetrable barriers 
in phase space''.} 
This invariant manifold is normally hyperbolic, i.e., it is of
saddle type, having \emph{stable and unstable manifolds} which are
of sufficient dimensionality to act as separatrices, i.e.  impenetrable
barriers to the dynamics which serve to partition energy surfaces
into regions of qualitatively distinct behaviour.

In addition, the NHIM forms the natural (dynamical) equator of a
(spherical) \emph{dividing surface} (DS) which locally divides the
energy surface into two components, one on either side of the
bottleneck.  The dividing surface is crossed (locally)
\emph{exactly once} by those trajectories which pass from one
component to the other; it is free of local recrossing by
trajectories and is of minimal directional flux, and is thus the
optimal dividing surface sought for in variational transition
state theory \cite{WaalkensWiggins04}.  In a nutshell, the
dividing surface enables a natural definition of the phase space
regions corresponding to \emph{reactants} and \emph{products}.  In
chemical terms, the equator of the dividing surface, the NHIM,
forms the energy surface of the \emph{activated complex},
consisting of an oscillating supermolecule poised between
the reactants and the products.

The purpose of this paper is to develop the geometrical intuition
associated with these high dimensional phase space structures that
govern reaction dynamics by constructing, step-by-step, three
representations  of the phase space geometry for $n=2$ DoF, and another model that 
illustrates various aspects of the phase space structures for an arbitrary number $n$ of DoF.
Moreover, we discuss the internal structure of the NHIM.  Before proceeding to the
develpment of these models we give a technical summary of the
relevant phase space structures, their interperation in terms of,
and implications for, reaction dynamics, and the method by which
they are realised (the Poincar\'e-Birkhoff normal form theory). 
All of the material described here can be found
in the above references.

%%%%%%%%%%%%%%%%%%%%%%%%%%%%%%%%%%%

\section{``Reaction-Type'' Dynamics and Phase Space Transition State Theory}%
\label{sec:summary_phases_space_structures}

We consider a Hamiltonian system with $n$ degrees of freedom, phase space coordinates $(Q,P)\in \R^n\times\R^n$ and Hamiltonian function $H$.
We assume that $(Q_0, P_0)$  is an equilibrium point of Hamilton's equations which is of saddle-centre-$\ldots$-centre stability type. 
\footnote{We will define this more precisely shortly. However, briefly, it means that the matrix associated with the linearization of Hamilton's equations about this equilibrium point has two real eigenvalues of equal magnitude, with one positive and one negative, and $n-1$ purely imaginary complex conjugate pairs of eigenvalues. We will assume that the eigenvalues satisfy a nonresonance condition that we will describe more fully in the following.}  
By adding a constant energy term which does not change the dynamics we can achieve 
 $H(Q_0,P_0)=0$, and for simplicity of exposition,
we can assume that the coordinates have been suitably
translated so that the relevant equilibrium point $(Q_0,P_0)$
is at the origin. For much of the discussion below, we will consider iso-energetic
geometrical structures belonging to a single positive energy surface
$\Sigma(h):=H^{-1}(h)$ for some constant $h>0$. 
In addition, we will typically restrict our attention to a certain
neighbourhood $\mathcal{L}$, local to the equilibrium point.  We
will defer until later a discussion of exactly how this region
is chosen; suffice it to say for now that the region is chosen so
that a new set of coordinates can be constructed (the normal form coordinates) in which the Hamiltonian can be expressed (the normal form Hamiltonian) such that it provides
an integrable nonlinear approximation to the dynamics which yields
phase space structures to within a given desired accuracy. 

Locally, the $(2n-1)$-dimensional energy surface $\Sigma(h)$ has
the structure of $\sphere^{2n-2}\times\mathbb{R}$ in the
$2n$-dimensional phase space.  The energy surface $\Sigma(h)$ is
split locally into two components, ``reactants''  and
``products'' , by a $(2n-2)$-dimensional  ``dividing surface''
that is diffeomorphic to $\sphere^{2n-2}$ and which we therefore denote by $\ts$.  The
dividing surface that we construct has the following properties:-

\begin{itemize}

\item The only way that trajectories can evolve from reactants  to products  (and
vice-versa), without leaving the local region $\mathcal{L}$, is
through $\ts$. In other words, initial conditions (ICs) on  this dividing surface  specify all reacting trajectories.

\item The dividing surface that we construct is free of local
recrossings; any trajectory which crosses it must leave the
neighbourhood $\mathcal{L}$ before it might possibly cross again.

\item The dividing surface that we construct minimizes the
 flux, i.e. the directional flux through the dividing surface will increase upon a generic deformation of the dividing surface (see \cite{WaalkensWiggins04} for the details).

\end{itemize}

The fundamental phase space building block that allows the construction of a dividing surface with these properties is a particular \emph{Normally Hyperbolic Invariant Manifold} (NHIM) which, for a
fixed positive energy $h$, will be denoted $\nhim$.  The
NHIM is diffeomorphic to $\sphere^{2n-3}$ and forms the natural
\emph{dynamical equator} of the dividing surface: The dividing
surface is split by this equator into $(2n-2)$-dimensional
hemispheres, each diffeomorphic to the open $(2n-2)$-ball,
$B^{2n-2}$.  We will denote these hemispheres by
$\tsf$ and $\tsb$ and call them the
``forward reactive'' and ``backward reactive'' hemispheres,
respectively.  $\tsf$ is crossed by trajectories
representing ``forward'' reactions (from reactants to products),
while $\tsb$ is crossed by trajectories representing
``backward'' reactions (from products to reactants).

The $(2n-3)$-dimensional NHIM can be viewed as the energy surface of an 
(unstable) invariant subsystem which as mentioned above, in
chemistry terminology, corresponds to the ``activated complex'', which as an
oscillating ``supermolecule''  is located between reactants and
products. 

The NHIM is of saddle stability type, having $(2n-2)$-dimensional
stable and unstable manifolds $W^s(h)$ and $W^u(h)$ that are
diffeomorphic to $\sphere^{2n-3}\times\mathbb{R}$.  Being of
co-dimension
\footnote{Briefly, the co-dimension of a submanifold is the dimension of the space in which the submanifold exists, minus the dimension of the submanifold. The significance of a submanifold being ``co-dimension one'' is that it is one less dimension than the space in which it exists. Therefore it can ``divide'' the space and act as a separatrix, or barrier, to transport.} 
one with respect to the energy surface, these invariant manifolds act as separatrices, partitioning the energy
surface into ``reacting'' and ``non-reacting'' parts as will explain in detail in Sec.~\ref{sec:structures_explicit}.

%-----------------------------------------------------------------------
\subsection{The normal form} \label{sec:normal_form}
%-----------------------------------------------------------------------

As mentioned in the previous section, reaction type dynamics are induced by
equilibrium points of saddle$\times$centre$\times\ldots\times$centre stability type. These are equilibria for which the matrix 
associated with the linearisation of Hamilton's equations
have eigenvalues which consist of a pair of real eigenvalues of
equal magnitude and opposite sign, $(+\lambda,-\lambda)$,
$\lambda\in\mathbb{R}$, and $(n-1)$ pairs of complex conjugate
purely imaginary eigenvalues, $(+i\omega_k,-i\omega_k)$,
$\omega_k\in\mathbb{R}$, for $k=2,\ldots,n$.

The phase space structures  near equilibria of this type exist independently of a specific coordinate system. However, in order to carry out specific calculations we will need to be able to express these phase space structures in coordinates. This is where Poincar\'e-Birkhoff normal form theory is used.This is a well-known theory and has been the subject of many
review papers and books, see,  e.g.,  \cite{Dep69,Meye74,DragtFinn76,AKN88,MeYe91,MeHa92,Murdock03}.  
For our purposes it provides an algorithm whereby the phase space structures described in the previous section can be realised
for a particular system by means of the normal form transformation
which involves making a nonlinear symplectic change of variables,

\begin{equation}\label{eq:NFtransform}
 (q, p) =T(Q, P),
\end{equation}

\noindent into \emph{normal form coordinates}, $(q,p)$ which, in a local
neighbourhood $\mathcal{L}$ of the equilibrium point, ``decouples'' the
dynamics into a ``reaction coordinate'' and ``bath modes.'' The coordinate transformation $T$ is obtained from imposing conditions on the form of $H$ expressed the new coordinates, $(q, p)$,

\begin{equation}
H_\text{NF}(q, p) = H\big(T^{-1}(q,p)\big)=H(Q,P).
\end{equation}

\noindent These conditions are chosen such that  $H_\text{NF}$ and the resulting equations of motions assume a simple form in which the reaction coordinate and bath modes ``decouple''.  
This decoupling  is one way of understanding how we are able to construct the phase space structures, in the normal form coordinates, that govern the dynamics of reaction. 

In fact, we will assume that a (generic)
non-resonance condition holds between the eigenvalues, namely that

\begin{equation}
k_2\omega_2+\dots+k_n\omega_n \ne 0 
\end{equation}

\noindent for all integer vectors $(k_2,\ldots,k_n)\in\mathbb{Z}^{n-1}$. 
\footnote{We note that the
inclusion of $\pm\lambda$ in a non-resonance condition would be
vacuous; one cannot have a resonance of this kind between a real
eigenvalue, $\pm\lambda$, and purely imaginary eigenvalues,
$\pm i\omega_k$, $k=2,\ldots,n$.}
When such a condition
holds, the NF procedure yields an explicit expression for the
normalised Hamiltonian $H_\text{NF}$ as a function of $n$ 
integrals of motion:

\begin{equation}
H_\text{NF}(q,p) = K_\text{NF}(I, J_2, \ldots, J_n)=\lambda I + \omega_2J_2+\ldots+\omega_nJ_n+\text{h.o.t.}.
\end{equation}
Here the higher order terms (h.o.t.) are at least of order two in the integrals $(I,J_2,\ldots,J_n)$.

\noindent The integral, $I$, corresponds to a ``reaction
coordinate'' (saddle-type DoF):

\begin{equation}\label{eq:def_I}
I = q_1p_1.
\end{equation}

\noindent We note that there is an equivalent form of the reaction
coordinate:- making the linear symplectic change of variables
$q_1=(\tilde{p}_1-\tilde{q})/\sqrt{2}$ and $p_1=(\tilde{p}_1+\tilde{q})/\sqrt{2}$, transforms the
above into the following form, which may be more familiar to many
readers,

\begin{equation}\label{eq:def_I_tilde}
I = q_1p_1 = \frac{1}{2}\left(\tilde{p}_1^2-\tilde{q}_1^2\right).
\end{equation}

\noindent Geometrically speaking, one can move freely between
these two representations by considering the plane $(q_1,p_1)$ and
rotating it by angle $\pi/4$, to give $(\tilde{q},\tilde{p})$.

The integrals $J_k$, for $k=2,\ldots,n$, correspond to ``bath
modes'' (centre-type \dof)\footnote{Throughout our work we use, somewhat interchangeably, terminology from both chemical reaction dynamics and dynamical systems theory. This is most noticable in our reference to the integrals of motion. $I$ is the integral related to reaction, and in the context of dynamical systems theory it is related to hyperbolic behaviour. The term ``reactive mode'' might also be used to describe the dynamics associated with this integral. The integrals $J_2, \ldots , J_n$ describe the dynamics associated with ``bath modes''. In the context of dynamical systems theory, the dynamics associated with these integrals is referred to as ``center type dynamics'' or ``center modes''. A key point here is that integrals of the motion provide us with the natural way of defining and describing the physical notion of a ``mode''. The nature of the mode is defined in the context of the specific application (i.e. chemical reactions) or, in the context of dynamical systems theory, through its stability properties (i.e. hyperbolic or centre).}:

\begin{equation}\label{eq:def_J}
J_k = \frac{1}{2}\left(p_k^2 + q_k^2\right),\quad k=2,\ldots,n\,.
\end{equation}

\noindent In the new coordinates, Hamilton's equations have a
particularly simple form:

\begin{equation} \label{eq:NF_equations_motion_gen}
\begin{split}
\dot{q}_1 = \hphantom{-}\frac{\partial{H_\text{NF}}}{\partial p_1}
&= \hphantom{-}\Lambda_1(I,J_2,\ldots,J_n)q_1,\\
\dot{p}_1 = {-}\frac{\partial{H_\text{NF}}}{\partial q_1}
&= {-}\Lambda(I,J_2,\ldots,J_n)p_1,\\
\dot{q}_k = \hphantom{-}\frac{\partial{H_\text{NF}}}{\partial p_k}
&= \hphantom{-}\Omega_k(I,J_2,\ldots,J_n)p_k\\
\dot{p}_k = {-}\frac{\partial{H_\text{NF}}}{\partial q_k} &=
{-}\Omega_k(I,J_2,\ldots,J_n)q_k,
\end{split}
\end{equation}

\noindent for $k=2,\ldots,n$, where we denote

\begin{eqnarray}
\Lambda(I,J_2,\ldots,J_n)  &:=&\frac{\partial{K_\text{NF}(I,J_2,\ldots,J_n)}}{\partial
I}\\
\Omega_k(I,J_2,\ldots,J_n) &:=&\frac{\partial{K_\text{NF}(I,J_2,\ldots,J_n)}}{\partial
J_k},\quad k=2,\ldots,n.
\end{eqnarray}

The integrals provide a natural definition of the term ``mode'' that is appropriate in the context of reaction, and they are a consequence of the (local) integrability in a neighborhood of the equilibrium point of saddle-centre-$\ldots$-centre stability type. Moreover, the expression of the normal form Hamiltonian in terms of the integrals provides us a way to partition the `energy' between the different modes. We will provide examples of how this can be done in the following.

The normal form transformation $T$ in \eqref{eq:NFtransform} can be computed in an algorithmic fashion. One can give explicit expression for the phase space structures discussed in the previous section in terms of the normal form coordinates, $(q,p)$. This way the phase space structures can be constructed in terms on the normal form coordinates, $(q,p)$, and for physical interpretation, transformed back to the orriginal ``physical'' coordinates, $(Q,P)$, by the inverse of the transformation $T$.\footnote{The original coordinates $(Q, P)$ typically have an interpretation as configuration space coordinates and momentum coordinates. The normal form coordinates $(q, p)$, in general, do not have such a physical interpretation since both $q$ and $p$ are nonlinear functions of both $Q$ and $P$.}.

In summary, the ``output'' of the normal form algorithm is the following:

\begin{itemize}

\item A symplectic transformation  $T(Q, P) = (q, p)$,  and its inverse $T^{-1}(q, p) = (Q, P)$, that relate the normal form coordinates $(q,p)$ to the original ``physical'' coordinates $(Q,P)$.

\item An expression for the normalized Hamiltonian: in the form, $H_\text{NF}$, in terms of the normal form coordinates $(q,p)$, and in the form $K_\text{NF}$, in terms of the integrals $(I,J_2,\ldots,J_n)$.

\item Explicit expressions for the integrals of motion $I$ and $J_k$, $k=2, \ldots, n$, in terms of the normal form coordinates.

\end{itemize}

%-----------------------------------------------------------------------
\subsection{Explicit definition of the phase space structures in the
normal form coordinates} \label{sec:structures_explicit}
%-----------------------------------------------------------------------

As indicated in the previous section it is straightforward to construct the local phase space objects
governing ``reaction'' in the NF coordinates, $(q, p)$.  In this
section, we will define the various structures in the normal form
coordinates and discuss briefly the consequences for the original
dynamical system.

\vspace*{0.3cm} \noindent {\bf The structure of an energy surface
near a saddle point: } For $h<0$, the energy surface consists of
two disjoint components.  The two components correspond to
``reactants'' and ``products.''  The top panel of
Fig.~\ref{fig:esurfs} shows how the two components project to the
various planes of the normal form coordinates.  The projection to
the plane of the saddle coordinates $(q_1,p_1)$ is bounded away
from the origin by the two branches of the hyperbola, $q_1p_1=I<0$, where $I$ is
given implicitly by the energy equation with the centre actions
$J_k$, $k=2,\dots,n$, set equal to zero:
$K_{\text{NF}}(I,0,\ldots,0)=h<0$.  The projections to the
planes of the centre coordinates, $(q_k,p_k)$, $k=2,\dots,n$, are
unbounded.

\def\figesurfs{%
Projection of energy surfaces (turquoise regions) to the planes of
the normal form coordinates.  The energy surface have energy 
$h<0$ (top panel),  $h=0$ (middle panel),  $h>0$ (bottom panel) . }
\def\FIGesurfs{
\centerline{\includegraphics[width=12.0cm]{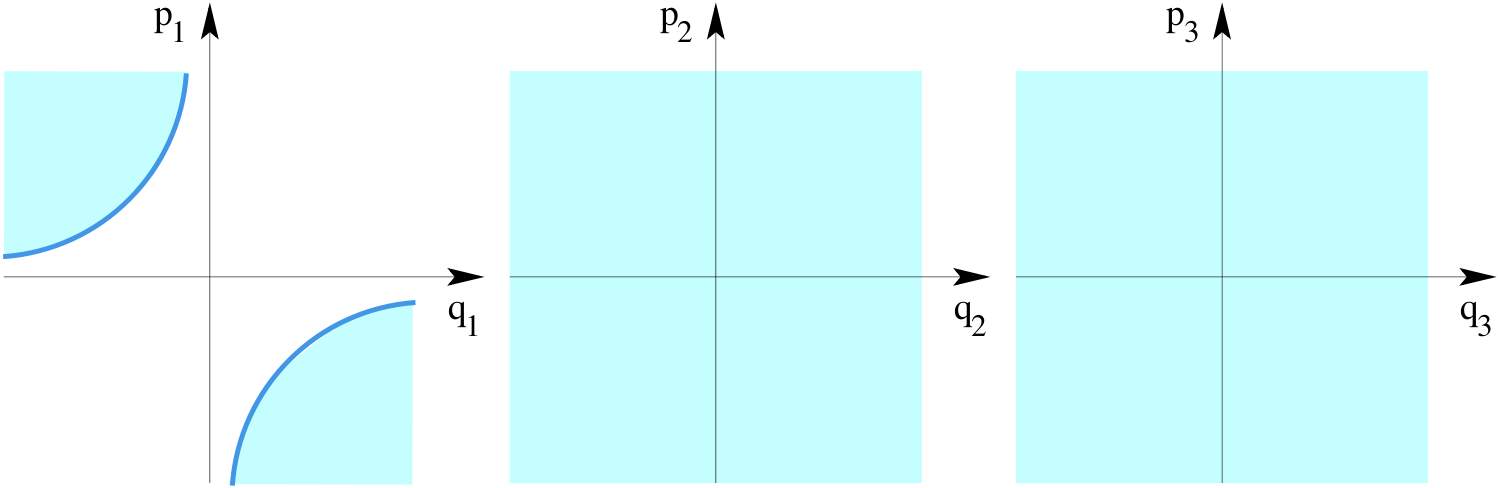}}
\centerline{\includegraphics[width=12.0cm]{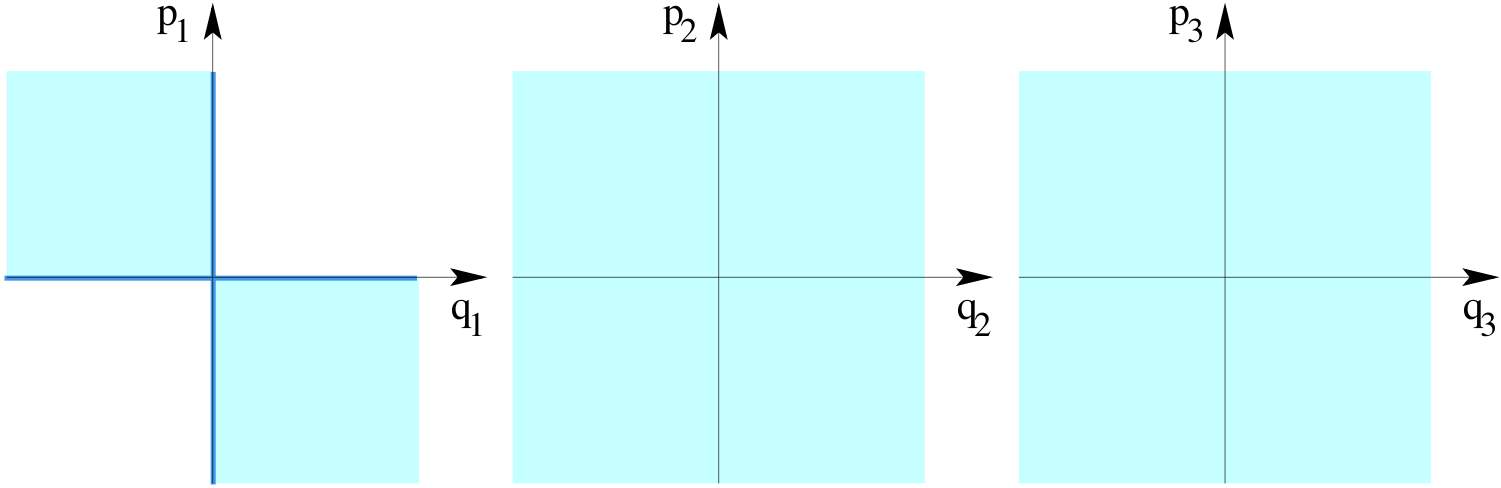}}
\centerline{\includegraphics[width=12.0cm]{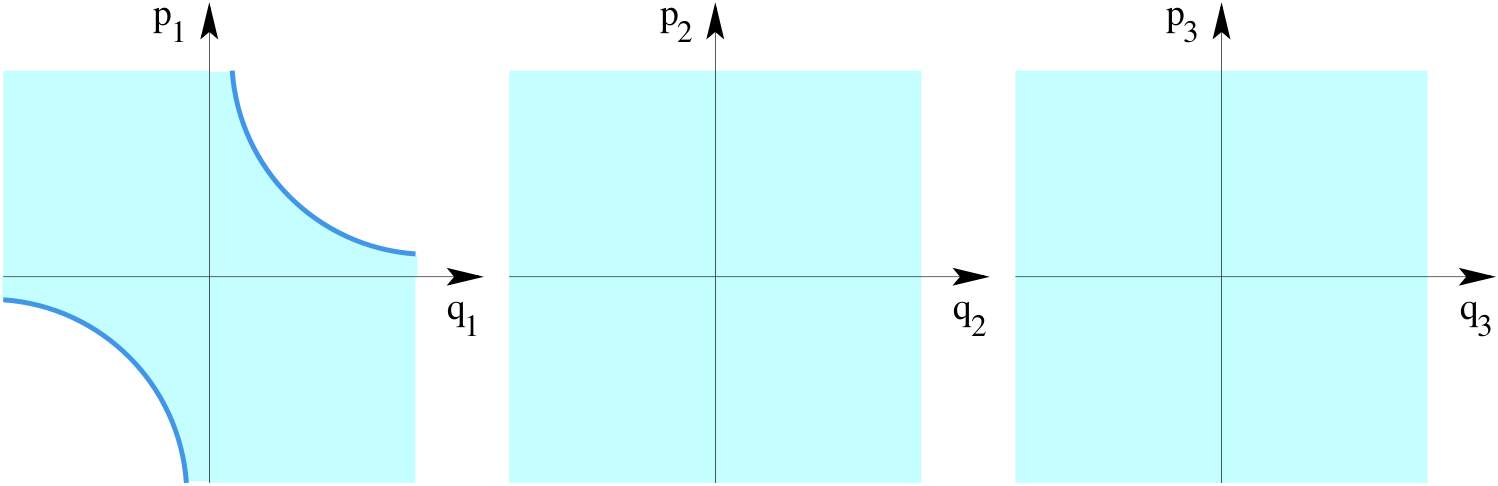}}
}
\FIGo{fig:esurfs}{\figesurfs}{\FIGesurfs}

At $h=0$, the formerly disconnected components merge (the energy
surface bifurcates), and for $h>0$, the energy surface has
locally the structure of a spherical cylinder, $\sphere^{2n-2}\times
\R$.  Its projection to the plane of the saddle coordinates now
includes the origin.  In the first and third quadrants it is
bounded by the two branches of the  hyperbola, $q_1p_1=I>0$, where $I$ is again given
implicitly by the energy equation with all centre actions equal to
zero, but now with an energy greater than $0$:
$K_{\text{NF}}(I,0,\dots,0)=h>0$. The projections to the planes
of the centre coordinates are again unbounded. This is illustrated
in the bottom panel of Fig.~\ref{fig:esurfs}.

\vspace*{0.3cm}
\noindent
{\bf The dividing surface, and reacting and nonreacting trajectories:}
On an energy surface with $h>0$, we define the dividing surface by
$q_1=p_1$.  This gives a $(2n-2)$-sphere which we denote by $\ts$.
Its projection to the saddle coordinates simply gives a line segment
through the origin which joins the boundaries of the projection of the
energy surface, as shown in Fig.~\ref{fig:tst}.  The projections of the
dividing surface to the planes of the centre coordinates are bounded by
circles $(p_k^2+q_k^2)/2=J_k$, $k=2,\dots,d$, where $J_k$ is
determined by the energy equation with the other centre actions, $J_l$,
$l\ne k$, and the saddle integral, $I$, set equal to zero.  The
dividing surface divides the energy surface into two halves,
$p_1-q_1>0$ and $p_1-q_1<0$, corresponding to reactants and products.

As mentioned above, trajectories project to
hyperbolae in the plane of the saddle coordinates, and to circles
in the planes of the centre coordinates.  The sign of $I$
determines whether a trajectory is nonreacting or reacting, see
Fig.~\ref{fig:tst}.  Trajectories which have $I<0$ are nonreactive and for one branch of the hyperbola $q_1p_1=I$ they stay on the
reactants side and for the other branch they stay on the products side; trajectories with $I>0$ are reactive, and for one  branch of the hyperbola $q_1p_1=I$ they react
in the forward direction, i.e., from reactants to products, and for the other branch they react in the
backward direction, i.e., from products to reactants.  The
projections of reactive trajectories to the planes of the centre
coordinates are always contained in the projections of the dividing
surface. In this, and other ways, the geometry of the reaction is
highly constrained.  There is no analogous restriction on the
projections of nonreactive trajectories to the centre coordinates.

\def\figtst{%
Projection of the dividing surface and reacting and nonreacting
trajectories to the planes of the normal form coordinates. In the
plane of the saddle coordinates, the projection of the dividing
surface is the  dark red diagonal line segment, which has
$q_1=p_1$.  In the planes of the centre coordinates, the
projections of the dividing surface are the dark red discs.
Forward and backward reactive trajectories (yellow and blue)
project to the first and third quadrant in the plane of the saddle
coordinates, respectively, and pass through the dividing surface.
The red and green curves mark nonreactive trajectories on the
reactant side ($p_1-q_1>0$), and on the product side
($p_1-q_1<0$), of the dividing surface, respectively.  The
turquoise regions indicate the projections of the energy surface.
}
\def\FIGtst{
\begin{center}
\includegraphics[width=12.0cm]{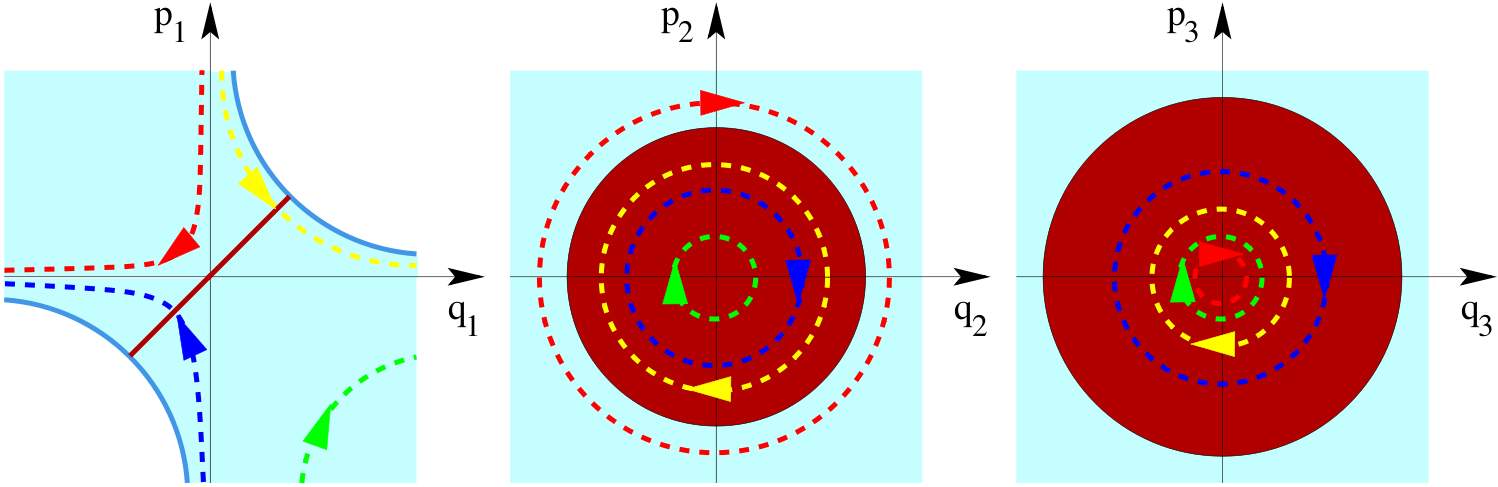}
\end{center}
}
\FIGo{fig:tst}{\figtst}{\FIGtst}

\vspace*{0.3cm} \noindent {\bf The normally hyperbolic invariant
manifold (NHIM) and its relation to the `activated complex':} On
an energy surface with $h>0$, the NHIM is given by $q_1=p_1=0$.
The NHIM has the structure of a $(2n-3)$-sphere, which we denote
by $\nhim$.  The NHIM is the equator of the dividing surface; it
divides it into two ``hemispheres'': the \emph{forward dividing
surface}, which has $q_1=p_1>0$, and the \emph{backward dividing
surface}, which has $q_1=p_1<0$.  The forward and backward
dividing surfaces have the structure of $(2n-2)$-dimensional
balls, which we denote by $\tsf$ and $\tsb$, respectively.  All
forward reactive trajectories cross $\tsf$; all backward reactive
trajectories cross $\tsb$.  Since $q_1=p_1=0$ in the equations of
motion \eqref{eq:NF_equations_motion_gen} implies that $\dot{q}_1=\dot{p}_1=0$, the
NHIM is an invariant manifold, i.e., trajectories started in the
NHIM stay in the NHIM for all time.  The system resulting from
$q_1=p_1=0$ is an invariant subsystem with one degree of freedom
less than the full system.  
In fact, $q_1=p_1=0$ 
defines the centre manifold associated with the saddle-centre-$\cdots$-centre equilibrium point, and 
the NHIM at an energy $h$ greater than the energy of the quilibrium point is given by the intersection of the centre manifold with the energy surface of this energy $h$ \cite{ujpyw,WaalkensWiggins04}.

This invariant subsystem with one degree of freedom less than the full system is the ``activated complex'' (in
phase space), located between reactants and products.  
The NHIM can be considered to
be the energy surface of the activated complex.  In particular,
all trajectories in the NHIM have $I=0$.

The equations of motion \eqref{eq:NF_equations_motion_gen} also show that
$\dot{p}_1-\dot{q}_1<0$ on the forward dividing surface $\tsf$, and
$\dot{p}_1-\dot{q}_1>0$ on the backward dividing surface $\tsb$.
Hence, except for the NHIM, which is is an invariant manifold, the
dividing surface is everywhere transverse to the Hamiltonian flow.
This means that a trajectory, after having crossed the forward or
backward dividing surface, $\tsf$ or $\tsb$, respectively, must leave
the neighbourhood of the dividing surface before it can possibly cross
it again. Indeed, such a trajectory must leave the local region in
which the normal form is valid before it can possibly cross the
dividing surface again.

The NHIM has a special structure: due to the conservation of the centre
actions, it is filled, or {\em foliated}, by invariant
$(n-1)$-dimensional tori, $\T^{n-1}$.  More precisely, for $d=3$ degrees of freedom, each
value of $J_2$ implicitly defines a value of $J_3$ by the energy
equation $K_{\text{NF}}(0,J_2,J_3)=h$.  For three degrees of freedom, the NHIM is thus
foliated by a one-parameter family of invariant 2-tori.  The end
points of the parameterization interval correspond to $J_2=0$ (implying
$q_2=p_2=0$) and $J_3=0$ (implying $q_3=p_3=0$), respectively.  At the
end points, the 2-tori thus degenerate to periodic orbits, the
so-called {\em Lyapunov periodic orbits}.  

\def\fignhim{%
The projection of the NHIM and the local parts of its stable
and unstable manifolds, $W^{\text{s}}(h)$ and $W^u(h)$, to the planes of the
normal form coordinates.  In the plane of the saddle coordinates,
the projection of the NHIM is the origin marked by the blue bold point, and the projection of
$W^\text{s}(h)$ and $W^u(h)$ are the $p_1$-axis and $q_1$-axis, respectively.
$W^\text{s}(h)$ consists of the forward and backward branches $W_{\text{f}}^{\text{s}}(h)$ and
$W_{\text{b}}^{\text{s}}(h)$, which have $p_1>0$ and $p_1<0$, respectively; $W^{\text{u}}(h)$
consists of $W_{\text{f}}^{\text{u}}(h)$ and $W_{\text{b}}^{\text{u}}(h)$, which have $q_1>0$ and $q_1<0$,
respectively.  In the plane of the centre coordinates, the projections of
the NHIM, $W^{\text{s}}(h)$, and $W^{\text{u}}(h)$ (the blue circular discs)
coincide with the projection of the dividing surface in
Fig.~\ref{fig:tst}.  The turquoise regions mark the projections
of the energy surface.
}
\def\FIGnhim{
\begin{center}
\includegraphics[width=12.0cm]{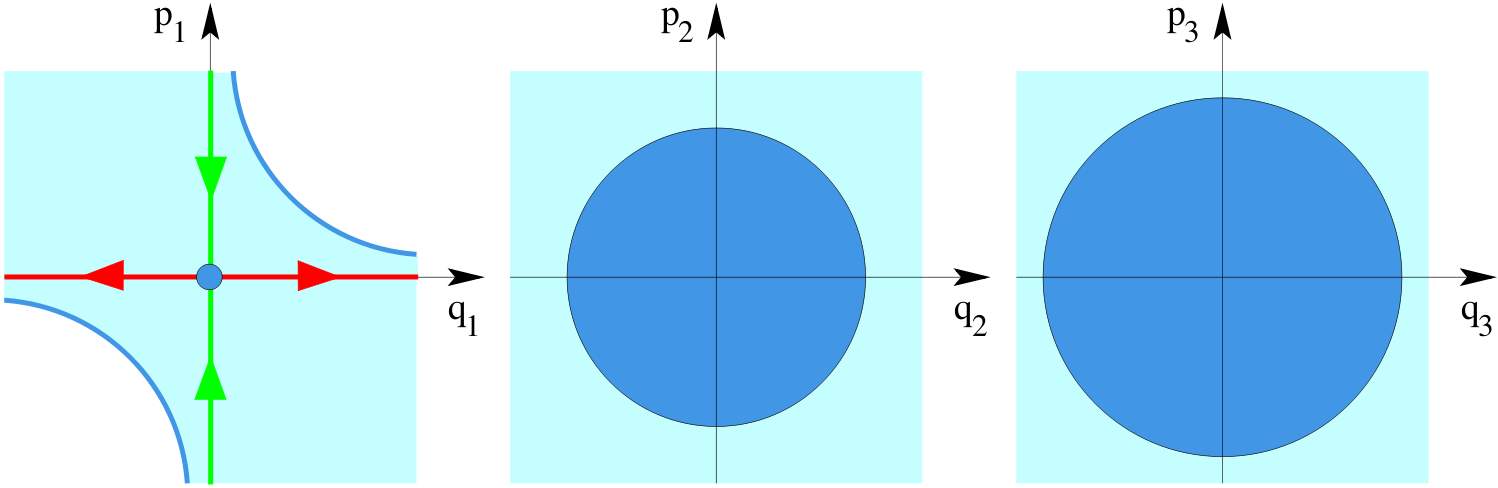}
\end{center}
}
\FIGo{fig:nhim}{\fignhim}{\FIGnhim}

\vspace*{0.3cm} \noindent {\bf The stable and unstable manifolds
of the NHIM forming the phase space conduits for reactions:} Since
the NHIM is of saddle stability type, it has stable and unstable
manifolds, $W^{\text{s}}(h)$ and $W^{\text{u}}(h)$.  The stable and unstable
manifolds have the structure of spherical cylinders,
$\sphere^{2n-3}\times\R$.  Each of them consists of two branches: the
``forward branches'', which we denote by $W_f^s(h)$ and
$W_{\text{f}}^{\text{u}}(h)$, and the ``backward branches'', which we denote by
$W_{\text{b}}^{\text{s}}(h)$ and $W_{\text{b}}^{\text{u}}(h)$.  In terms of the normal form
coordinates, $W_{\text{f}}^{\text{s}}(h)$ is given by $q_1=0$ with $p_1>0$,
$W_{\text{f}}^{\text{u}}(h)$ is given by $p_1=0$ with $q_1>0$, $W_b^s(h)$ is given
by $q_1=0$ with $p_1<0$, and $W_b^u(h)$ is given by $p_1=0$ with
$q_1<0$, see Fig.~\ref{fig:nhim}.  Trajectories on these manifolds
have $I=0$.

Since the stable and unstable manifolds of the NHIM are of one less
dimension than the energy surface, they enclose volumes of the energy
surface.  We call the union of the forward branches, $W_f^s(h)$ and
$W_{\text{f}}^{\text{u}}(h)$, the {\em forward reactive spherical cylinder} and denote it by
$W_{\text{f}}(h)$.  Similarly, we define the {\em backward reactive spherical
cylinder}, $W_{\text{b}}(h)$, as the union of the backward branches, $W_{\text{b}}^{\text{s}}(h)$ and
$W_{\text{b}}^{\text{u}}(h)$.

\def\figvolumes{%
Projections of the reactive volumes enclosed by the forward
and backward reactive spherical cylinders, $W_{\text{f}}(h)$ and $W_{\text{b}}(h)$, and
the forward and backward reactions paths, to the planes of the normal
form coordinates.  The volumes enclosed by $W_{\text{f}}(h)$ and $W_{\text{b}}(h)$
project to the dark pink and green regions in the first and third quadrant
in the plane of the saddle coordinates, respectively.
These volumes project to the dark green/dark pink brindled disks in the planes of the centre
coordinates, where their
projections coincide with the
projection of the NHIM and the dividing surface in
Figs.~\ref{fig:tst}~and~\ref{fig:nhim}.  The forward and backward
reaction paths project to the two branches of a hyperbola marked blue in the first and
third quadrant in the plane of the saddle coordinates, respectively,
and to the origins (bold blue points) in the planes of the centre
coordinates.  The turquoise regions mark the projections of the
energy surface.
}
\def\FIGvolumes{
\begin{center}
\includegraphics[width=12.0cm]{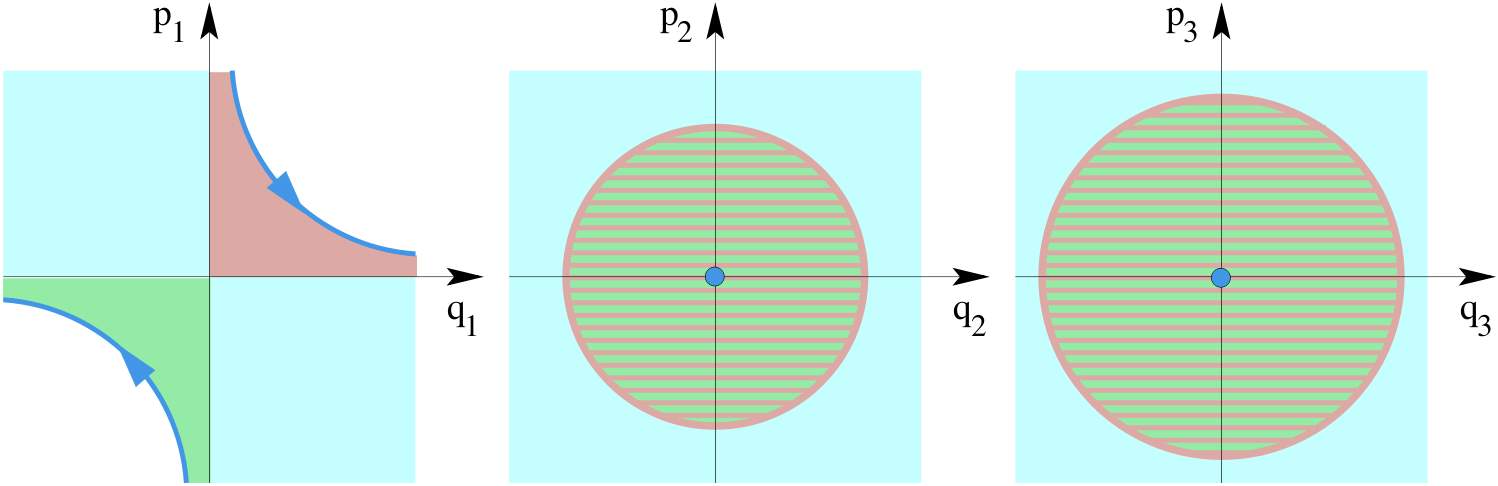}
\end{center}
}
\FIGo{fig:volumes}{\figvolumes}{\FIGvolumes}

The reactive volumes enclosed by $W_f(h)$ and $W_b(h)$ are shown
in Fig.~\ref{fig:volumes} as their projections to the normal form
coordinate planes.  In the plane of the saddle coordinates, the
reactive volume enclosed by $W_f(h)$ projects to the first
quadrant.  This projection is bounded by the corresponding
hyperbola $q_1 p_1= I$, with $I$ obtained from
$K_{\text{NF}}(I,0,\dots,0)=h$.  Likewise, $W_b(h)$ projects to
the third quadrant in the $(q_1,p_1)$-plane.  $W_f(h)$ encloses
{\em all} forward reactive trajectories; $W_b(h)$ encloses {\em
all} backward reactive trajectories.  {\em All} nonreactive
trajectories are contained in the complement.

\vspace*{0.3cm}
\noindent
{\bf Forward and backward reaction paths:}
The local geometry of $W_{\text{f}}(h)$ and $W_{\text{b}}(h)$ suggests a natural definition of
{\em dynamical} forward and backward reaction paths as the unique
paths in {\em phase space} obtained by putting all of the energy of a
reacting trajectory into the reacting mode, i.e., setting
$q_2=\dots=q_n=p_2=\dots=p_n=0$.  This gives the two branches of the  hyperbola $q_1
p_1 = I$, with $I$ obtained from $K_{\text{NF}}(I,0,\dots,0)=h$, which in phase space are contained in the plane of
the saddle coordinates, see Fig.~\ref{fig:volumes}.  This way, the
forward (respectively, backward) reaction path can be thought of as
the ``centre curve'' of the relevant volume enclosed by the forward
(resp., backward) reactive spherical cylinder $W_{\text{f}}(h)$ (resp.,
$W_{\text{b}}(h)$).  These reaction paths are the special reactive trajectories
which intersect the dividing surface at the ``poles'' (in the sense of
North and South poles, where $q_1=p_1$ assumes its maximum and minimum
value on the dividing surface).

\vspace*{0.3cm} \noindent {\bf The Transmission Time through the
Transition State Region:} The normal form coordinates provide a way
of computing the time for {\em all} trajectories to cross the
transition region. We illustrate this with a forward reacting
trajectory (a similar argument and calculation can be applied to
backward reacting trajectories). We choose the boundary for the
entrance to the reaction region to be $p_1-q_1=c$ for some
constant $c>0$, i.e., initial conditions which lie on the reactant
side of the transition state, and the boundary for exiting the
reaction region to be $p_1-q_1=-c$ on the product side. We now
compute the {\em time of flight} for a forward reacting trajectory
with initial condition on $p_1-q_1=c$ to reach $p_1-q_1=-c$ on the
product side.  The solutions are
$q_1(t)=q_1(0) \exp (
\Lambda (I,J_2,\dots,J_n) t )$ and
$p_1(t)=p_1(0) \exp ( - \Lambda (I,J_2,\dots,J_n) t)$ (see \eqref{eq:NF_equations_motion_gen}),
where $\Lambda (I,J_2,\dots,J_n)$ is determined by the
initial conditions.  This gives the time of flight as

\begin{equation}
T = \left( \Lambda(I,J_2,\dots,J_n) \right)^{-1}
 \ln \left( \frac{p_1(0)}{q_1(0)} \right).
\end{equation}

\noindent
 The time diverges logarithmically as $q_1(0)\rightarrow
0$, i.e., the closer the trajectory starts to the boundary
$W_{\text{f}}(h)$. It is not difficult to see that the time of flight is
shortest for the centre curve of the volume enclosed by $W_{\text{f}}(h)$,
i.e., {\em the trajectory which traverses the transition state
region fastest is precisely our forward reaction path}.  A similar
construction applies to backward reactive trajectories.

In fact, the normal
form can be used to map trajectories through the transition state
region, i.e. the phase space point at which a trajectory enters
the transition state region can be mapped analytically to the
phase space point at which the trajectories exits the transition
state region.

%%%%%%%%%%%%%%%%%%%%%%%%%%%%%%%%%%%%%%%%%%%%%

\subsection{The foliation of the reaction region by Lagrangian submanifolds}
\label{sec:Lagrange_foliation}

The existence of the $n$ integrals of motion, $(I,J_2,\ldots, J_n)$, induce phase space structures which lead to further constraints on the trajectories in addition to the ones described above. In order to describe these structures and the resulting constraints it is useful to introduce the so called \emph{momentum map}, ${\cal M}$ \cite{Gui94,MR99}
which maps a point
$(q_1,\ldots,q_n,p_1,\ldots,p_n)$ in the phase space $\R^{n}\times\R^n$ to
the $n$ integrals evaluated at this point:
\begin{equation} \label{eq:def_momentum_map}
{\cal M}:\R^n\times\R^n \to \R^n\,,\quad(q_1,\ldots,q_n,p_1,\ldots,p_n) \mapsto (I,J_2,\ldots,J_n)\,.
\end{equation}
The preimage of a value for the constants of motion $(I,J_2,\ldots,J_n)$ under ${\cal M}$  is called a \emph{fibre}. A fibre thus corresponds to the common level set of the integrals in phase space.

A point $(q_1,\ldots,q_n,p_1,\ldots,p_n)$ is called a \emph{regular
point} of the momentum map if the linearisation of the momentum map, D$\M$, has full rank $n$
at this point, i.e., if the gradients of the $n$ integrals $I$, $J_k$,
$k=2,\ldots,n$, with respect to the
phase space coordinates $(q,p)$ are linearly independent at this point. If the rank of D$\M$ is less than $n$ then the point is called irregular.
A fibre is called regular if it
consists of regular points only. Else it is called an irregular fibre. 
In fact almost all fibres are regular.
They  are $n$-dimensional
manifolds given by the Cartesian product of an hyperbola $q_1 p_1 =
I$ in the saddle plane $(q_1,p_1)$ and $n-1$ circles
$\sphere^1$ in the centre planes $(q_k,p_k)$, $k=2,\ldots,n$.
Since the hyperbola $q_1 p_1 =
I$ consists of two branches
each of which have the topology of a line $\R$,  the
regular fibres consist of two disjoint \emph{toroidal cylinders}, $\T^{n-1}
\times \R $,
which are the Cartesian products of a
$(n-1)$-dimensional torus and a line.
We denote these toroidal cylinders by
\begin{equation} \label{eq:def_Lambda_plus}
\Lambda^+_{I,J_2,\ldots,J_n} =
\{(q,p)\in \R^{2n}\,:\,
  p_1q_1=I,\,\frac12\big(p_2^2+q_2^2\big)=J_2\,,\ldots\,,\frac12\big(p_n^2+q_n^2\big)=J_n\,,q_1>0 \}
\end{equation}
and
\begin{equation}\label{eq:def_Lambda_minus}
\Lambda^-_{I,J_2,\ldots,J_n} =
\{(q,p)\in \R^{2n}\,:\,
  p_1q_1=I,\,\frac12\big(p_2^2+q_2^2\big)=J_2\,,\ldots\,,\frac12\big(p_n^2+q_n^2\big)=J_n\,,q_1<0 \}\,.
\end{equation}
$\Lambda^+_{I,J_2,\ldots,J_d}$ and $\Lambda^-_{I,J_2,\ldots,J_d}$ are
\emph{Lagrangian manifolds} \cite{Arnold78}. The Lagrangian manifolds consists of all trajectories which have the same constants of motion. In particular the Lagrangian manifolds are invariant, i.e. a trajectory with initial condition on a Lagrangian manifold will stay in the Lagrangian manifold for all time. 
For $I<0$, the Lagrangian manifolds $\Lambda^-_{I,J_2,\ldots,J_n}$ and $\Lambda^+_{I,J_2,\ldots,J_n}$
consist of nonreactive trajectories in the reactants resp. products components of the energy surface. For $I>0$, $\Lambda^+_{I,J_2,\ldots,J_n}$ consists of forward reactive trajectories, and $\Lambda^-_{I,J_2,\ldots,J_n}$ consists of backward reactive trajectories.

We will discuss the irregular fibres in more detail in  Sec.~\ref{sec:ndof}.

%%%%%%%%%%%%%%%%%%%%%%%%%%%%%%%%%%%%%%%%%%%%%%%%%%%%

\subsection{Implications for the original system}

The normalization procedure proceeds via
formal power series manipulations whose input is a Taylor
expansion of the original Hamiltonian, $H$, necessarily up to some
finite order, $M$, in homogeneous polynomials.  For a particular
application, this procedure naturally necessitates a suitable
choice of the order, $M$, for the normalization, after which one
must make a restriction to some local region, $\mathcal{L}$, about
the equilibrium point in which the resulting computations achieve
some desired accuracy. Hence, the accuracy of the normal form as a power series expansion truncated
at order $M$ in a neighborhood $\mathcal{L}$ is determined by comparing the dynamics associated with the normal form with the dynamics of the original system. There are several independent tests that can be carried out to verify accuracy of the normal form. Straightforward tests that we use are the following:

\begin{itemize}

\item Examine how the integrals associated with the normal form change on trajectories of the full Hamiltonian (the integrals will be constant on trajectories of the normal form). 

\item Check invariance of the different invariant manifolds (i.e. the NHIM and its stable and unstable manifolds) with respect to trajectories of the full Hamiltonian.

\end{itemize}

Both of these tests will require us to use the transformations between the original coordinates and the normal form coordinates. Specific examples where $M$, $\mathcal{L}$ and accuracy of the normal forms are considered can be found in \cite{WaalkensBurbanksWiggins04,WaalkensBurbanksWigginsb04,WaalkensBurbanksWiggins05,WaalkensBurbanksWiggins05b,WaalkensBurbanksWiggins05c}.

%%%%%%%%
%%%%%%%%
%\input{model_I}
%%%%%%%%
%%%%%%%%

%-----------------------------------------------------------------------
\section{Models of the local phase space structures for 2-DoF systems}
%-----------------------------------------------------------------------
\label{sec:2dofModels}

In the discussion of the phase space structures governing reaction type dynamics in Sec.~\ref{sec:structures_explicit} we visualized these structures by projecting them to the 2-dimensional planes of the coordinates and conjugate momenta of the normal form. The topology of the phase space structures and the interrelationship between them is to a large extent obscured in these projection because of the lacking injectivity of the projections. For $n=2$ \dof however, a surface of constant energy  containing the  phase space structures is three-dimensional. For this case, 
it is therefore possible to visualize the phase space structures in three-dimensional space. In this section we will develop three models to explicitly implement such  visualizations. Our discussion will be in the context of linear Hamiltonian systems since they allow us to describe the phase space structures in terms of explicit formulae. The influence of nonlinear terms is discussed   
in Sec.~\ref{sec:nonlinear_case}.

Therefore we begin  by considering the quadratic Hamiltonian 

\begin{equation}
H(x,y,p_x,p_y) = \frac{\lambda}{2}\big( p_x^2 -x^2 \big) + \frac{\omega}{2}\big( p_y^2+ y^2  \big)
\label{eq:ham2dof}
\end{equation}
with $\lambda,\omega>0$.

\noindent This gives the linear Hamiltonian vector field

\begin{eqnarray}\label{eq:2Dlinear}
\dot{x} & = & \phantom{-} \frac{\partial H}{\partial p_x} = \lambda p_x, \label{x_comp} \\
\dot{p}_x & = & -\frac{\partial H}{\partial x}= \lambda x, \label{px_comp} \\
\dot{y} & = & \phantom{-} \frac{\partial H}{\partial p_y} =  \omega p_y, \label{y_comp} \\
\dot{p}_y & = & -\frac{\partial H}{\partial y}= - \omega y,\label{py_comp} \,.
\end{eqnarray}

The Hamiltonian \eqref{eq:ham2dof} is already in normal form. 
To keep the notation simple we will not specifically indicate this by a subscript NF. In fact we have
\begin{equation} \label{eq:def_K_2dof}
H(x,y,p_x,p_y) = K(I,J) =  \lambda I + \omega J \,,
\end{equation}
where $I$ and $J$ are the integrals given in \eqref{eq:def_I_tilde}) and 
\eqref{eq:def_J}, respectively.

\noindent For a fixed energy $h$, the three dimensional energy surface is given by the level set of 
the Hamiltonian (\ref{eq:ham2dof})   as follows:

\begin{equation} \label{eq:energy2dof}
\Sigma(h) = \{  (x,y,p_x,p_y) \in \R^4 \,:\, \lambda\big( p_x^2- x^2 \big)+ \omega\big(  p_y^2 + y^2 \big) =2h\}\,.
\end{equation}

\noindent
We rewrite the energy equation in  the form

\begin{equation}\label{eq:energy2dof_x}
\lambda p_x^2 + \omega p_y^2 +  \omega y^2 =  2h + \lambda x^2\,,
\end{equation}

\noindent
and fix $x$ such that the right hand side of this equation is positive. 
Note that if $h>0$  the right hand side  will be positive for any fixed $x\in \R$, while if $h<0$ it  
will be positive only for $|x|>\sqrt{-2h/\lambda}$, i.e. for $x$  sufficiently small or large.
For such an $x$,  Eq.~(\ref{eq:energy2dof_x}) defines an ellipsoid 
 in the three-dimensional Euclidean  space,  $\R^3$, with coordinates  $( p_x,p_y,y)$. 
 This ellipsoid is rotationally symmetric about the $p_x$-axis. It 
has two  semi-axes of length
$\big((2h+\lambda x^2)/\omega\big)^{1/2}$ and one semi-axis of length $\big((2h+\lambda x^2)/\lambda\big)^{1/2}$.  
Topologically, this is a 
two-dimensional sphere, $\sphere^2$,  which we denote by $\sphere_{x}^2(h)$, i.e.

\begin{equation}\label{eq:def_sphere_x_h}
   \sphere^2_{x}(h) = \{(p_x,p_y,y) \in \R^3 \,:\,   \lambda p_x^2 + \omega\big( p_y^2 + y^2 \big) =  2h + \lambda x^2 \}\,.
\end{equation}

\noindent
The energy surface thus appears to have the structure of a one-parameter family  of two-dimensional spheres parametrized by $x$, where the `radius'  (the length of the shortest semi-axis of the ellipsoid \eqref{eq:energy2dof_x}) can shrink to zero if $h<0$. 
The three models of the energy surface and the phase space structures contained in it that we present in the following differ by the way of representing the  2-spheres $\sphere^2_{x}(h)$. For clarity, let us summarize the phase space structures discussed in 
Sec.~\ref{sec:summary_phases_space_structures} for the Hamiltonian \eqref{eq:ham2dof} before we proceed. We will restrict ourselves to the case $h>0$.

The dividing surface is obtained by setting $x=0$ on the energy surface. This gives the two-dimensional sphere

\begin{equation} \label{eq:def_ds_2d}
\tstwoDOF =   \Sigma(h) \cap \{ x=0 \}  =  \{ (x,y,p_x,p_y) \in \R^4  \,:\,      \lambda p_x^2 + \omega\big(  p_y^2 + y^2 \big) =2h, \, x=0 \}\,.
\end{equation}

\noindent
For the 2-\dof\  case, the NHIM is an unstable periodic orbit (the Lyapunov periodic orbit associated with the saddle point). For the Hamiltonian  \eqref{eq:ham2dof}, it is obtained by setting $x=0$ and $p_x=0$ on the energy surface. This gives the circle

\begin{equation}
\po =
 \Sigma(h) \cap \{ x=0,\, p_x=0  \}  =  \{ (x,y,p_x,p_y) \in \R^4  \,:\,     \omega\big(  p_y^2 + y^2 \big) =2h,\, x=0,\, p_x=0  \}\,.
\end{equation}

\noindent
The NHIM divides the dividing surface, $\tstwoDOF$, into the forward hemisphere

\begin{equation}
\tsftwoDOF =   \Sigma(h) \cap \{ x=0,\,p_x>0  \}  =  \{ (x,y,p_x,p_y) \in \R^4  \,:\,     p_x= 
 \sqrt{\big( 2h- \omega\big(  p_y^2 + y^2 \big)  \big)/\lambda  },\,x=0 \}
\end{equation}

\noindent
and the
backward hemisphere

\begin{equation}
\tsbtwoDOF =   \Sigma(h) \cap \{ x=0,\, p_x<0  \}  =  \{ (x,y,p_x,p_y) \in \R^4  \,:\,      p_x= 
- \sqrt{\big( 2h- \omega\big(  p_y^2 + y^2 \big)  \big)/\lambda  },\,x=0 \}\,,
\end{equation}

\noindent
which are two dimensional balls or disks. 
The stable manifold $W^s$ of $\po $ consists of the
forward branch

\begin{equation}
W^s_{\text{f}} =  \Sigma(h) \cap \{ x = -p_x,\, x<0  \} =   
\{ (x,y,p_x,p_y) \in \R^4  \,:\,       \omega\big(  p_y^2 + y^2 \big) =2h,\, x=-p_x,\,x<0  \}
\end{equation}

\noindent
and the backward branch

\begin{equation}
W^s_{\text{b}} =  \Sigma(h) \cap \{ x = -p_x,\, x>0  \} =   
\{ (x,y,p_x,p_y) \in \R^4  \,:\,       \omega\big(  p_y^2 + y^2 \big) =2h,\, x=-p_x,\,x>0  \}\,.
\end{equation}

\noindent
Similarly, the unstable manifold $W^u$ of $\po $ consists of the
forward branch

\begin{equation}
W^u_{\text{f}} =  \Sigma(h) \cap \{ x = p_x,\, x>0  \} =   
\{ (x,y,p_x,p_y) \in \R^4  \,:\,       \omega\big(  p_y^2 + y^2 \big) =2h,\, x=p_x,\,x>0  \}
\end{equation}

\noindent
and the backward branch

\begin{equation}
W^u_{\text{b}} =  \Sigma(h) \cap \{ x = p_x,\, x<0  \} =   
\{ (x,y,p_x,p_y) \in \R^4  \,:\,       \omega\big(  p_y^2 + y^2 \big) =2h,\, x=p_x,\,x<0  \}\,.
\end{equation}

\noindent
Finally, the forward and backward reaction paths are obtained from putting all the energy in the $x$ \dof. This gives the lines

\begin{equation}
 \Sigma(h) \cap \{ y = p_y=0,\, p_x>0  \} = 
 \{ (x,y,p_x,p_y) \in \R^4  \,:\,      p_x = \sqrt{(2h +\lambda x)/\lambda } , \, y=p_y=0 \} 
\end{equation}

\noindent
for the forward reaction path, and

\begin{equation} \label{eq:def_forward_reaction_path_2D}
 \Sigma(h) \cap \{ y = p_y=0,\, p_x<0  \} = 
 \{ (x,y,p_x,p_y) \in \R^4  \,:\,      p_x = -\sqrt{(2h +\lambda x)/\lambda } , \, y=p_y=0 \} 
\end{equation}

\noindent
for the backward reaction path.

Note that  the energy surface \eqref{eq:energy2dof}, the phase space structures \eqref{eq:def_ds_2d}-\eqref{eq:def_forward_reaction_path_2D} and also the Lagrangian cylinders defined in \eqref{eq:def_Lambda_plus} and \eqref{eq:def_Lambda_minus} are invariant under  the $\sphere^1$ symmetry action 
\begin{equation} \label{eq:S1_symmetry}
(x,y,p_x,p_y)\to(x,\ue^{-\ui \varphi } y,p_x, \ue^{-\ui \varphi}  p_y)\,,\quad \varphi \in \sphere^1 \,.
\end{equation}  

%-----------------------------------------------------------------------
\subsection{Model I: projection of 2-spheres to their equatorial planes}
\label{sec:model-hw}
%-----------------------------------------------------------------------

The main idea of the first model is to view the 2-sphere \eqref{eq:def_sphere_x_h} as consisting of two hemispheres and an equator, i.e. as the disjoint union of two open two-dimensional disks or balls, $B^2$, and a circle, or equivalently one-dimensional sphere, 
$\sphere^1$, and using parametrizations of these geometric structures to parametrize the energy surface. 

\begin{figure}[htb!]
\begin{center}
\includegraphics[width=2.9in]{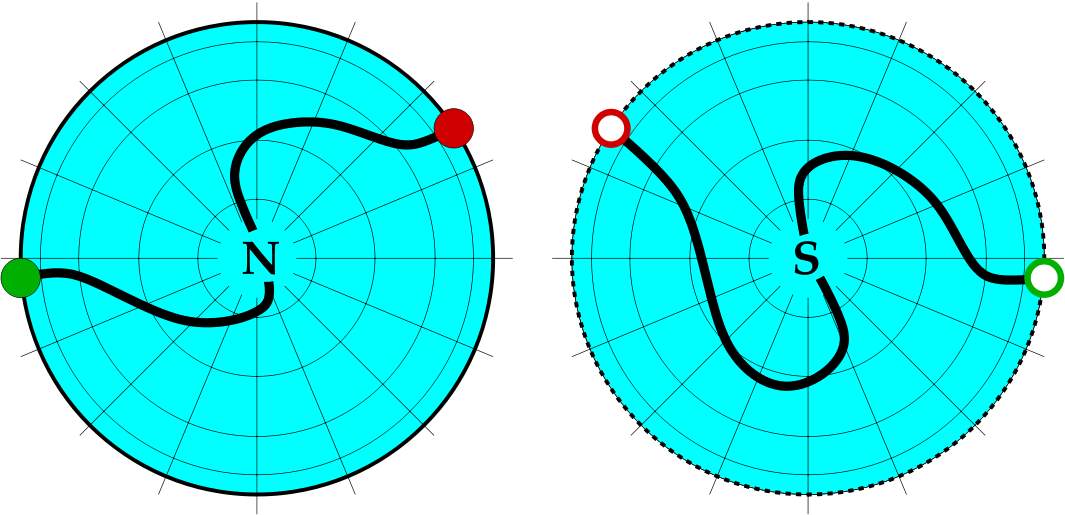}
\caption{\label{fig:nhim-s2-model}  A model of the sphere $\sphere^2$
formed by taking two open discs ($2$-balls) $B^2_\text{North}$ and
$B^2_\text{South}$ and identifying points on their boundaries; the
``equator'' $\sphere^1_\text{Equator}$. Also shown is the representation of a
smooth closed curve on $\sphere^2$ which passes through the North and South
poles, $N$ and $S$; the points at which the representation of the
curve ``jumps'' from one disc in the model to the other disc are
marked by dots.}
\end{center}
\end{figure}

The decomposition of a 2-sphere, $\sphere^2$, in terms of 2-balls and a circle is familiar from visualizing the
surface of the Earth  by drawing two
filled-in disks ($B^2$), one each for the Northern and Southern
hemispheres, and identifying the boundaries of the two disks,

\begin{equation} \label{eq:decomp_S_2}
\sphere^2 \equiv B^2_{\text{North}} \cup \sphere^1_{\text{Equator}} \cup
B^2_{\text{South}},
\end{equation}

\noindent
which we illustrate in Fig.~\ref{fig:nhim-s2-model}.

For the 2-spheres  $\sphere^2_{x}(h)$  in Eq.~\eqref{eq:def_sphere_x_h}, we define the decomposition \eqref{eq:decomp_S_2} as 

\begin{equation}
   \sphere^2_{x}(h) = B^2_{+,x}(h) \cup  \sphere^1_{0,x}(h) \cup B^2_{-,x}(h)\,,
\end{equation}

\noindent
where

\begin{eqnarray}  
  B^2_{+,x}(h) &=& \{ (p_x,p_y,y) \in \R^3  \,:\, \omega \big( p_y^2 + y^2\big) <2h- \lambda x^2,  \, p_x = \phantom{-} \sqrt{(2h + \lambda x^2 - \omega p_y^2 - \omega y^2)/\lambda}   	> 0  \} \,, \label{eq:defB2plus} \\
  B^2_{-,x}(h) &=& \{ (p_x,p_y,y) \in \R^3  \,:\, 
 \omega \big( p_y^2 + y^2\big) <2h- \lambda x^2 , \, 
  p_x = -\sqrt{(2h + \lambda x^2 - \omega p_y^2 - \omega y^2)/\lambda}   < 0  \} \,, \label{eq:defB2minus} \\
  \sphere^1_{0,x}(h) &=& \{ (p_x,p_y,y) \in \R^3  \,:\, 
\omega \big( p_y^2 + y^2\big) =2h- \lambda x^2 , \, 
  p_x =  0  \}\,. \label{eq:defequator} 
\end{eqnarray}

\noindent
We can view the manifolds  $B^2_{+,x}(h) $ and $B^2_{-,x}(h) $
as  graphs of the functions

\begin{equation} \label{eq:fpm}
\ccp_{\pm,x} : \tilde{B}^2_{x} (h) \rightarrow \R\,,\quad (y,p_y)  \mapsto \pm \sqrt{(2h + \lambda x^2 -\omega p_y^2 - \omega y^2)/\lambda}  \, ,
\end{equation}

\noindent
where the domains are the circular disks

\begin{equation} \label{eq:plane_2_balls}
\tilde{B}^2_{x}(h) = \{ (p_y,y) \in \R^2  \,:\, \omega \big(p_y^2 + y^2 \big)  < 2h + \lambda x^2    \}  \subset \R^2 \,.
\end{equation}

\begin{figure}[!htb]
\begin{center}
  \includegraphics[width=12cm]{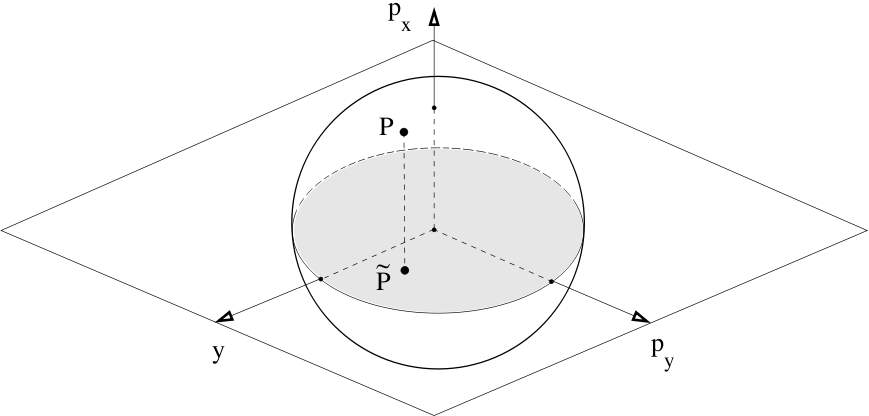}     
\end{center}
   \caption[]{\label{fig:projection_standard} Model I: projection of $\sphere^2_{x}(h)$ to the $(y,p_y)$-plane.}
\end{figure}

\noindent
The graphs `join' at the equator 
$ \sphere^1_{0,x}(h)$.  In order to geometrically visualize $\sphere^2_{x}(h)$ we take two copies of the circular disks 
$\tilde{B}^2_{x}(h)$ in the $(y, p_y)$-plane where $p_x$ is given by the function $\ccp_{+,x}$  on one copy which we denote by 
$\tilde{B}^2_{+,x}(h) $ and the function $\ccp_{-,x}$ on the other copy which we denote by  $\tilde{B}^2_{-,x}(h) $.
$\tilde{B}^2_{\pm,x}(h)$ thus denote the images of   $B^2_{\pm,x}(h)$ under the projection

\begin{equation}\label{eq:def_pi_p_x}
\pi_{p_x\,x}:  B^2_{\pm,x}(h) \to \R^2 \,, \quad  (p_x,p_y,y) \mapsto (y,p_y)\,,   
\end{equation}

\noindent
see Fig.~\ref{fig:projection_standard}.
Along the boundaries  of $\tilde{B}^2_{+,x}(h)$ and $\tilde{B}^2_{-,x}(h)$ we have  $p_x=0$, and the points on the boundaries  
which have the same $(y,p_y)$ are identified. 

The first model (Model I) of the energy surface and the phase space structures contained in it is obtained
from defining a map from the energy surface $\Sigma(h)$ to 
$\R^3$ using the projections $\pi_{p_x\,x}$, $x\in \R$,  in \eqref{eq:def_pi_p_x} to map points $(x,y,p_x,p_y) \in \Sigma(h)$ according to

\begin{equation}
F^{\text{I}}   : 
(x,y,p_x,p_y) \mapsto  (x,\pi_{p_x\,x} (p_x,p_y,y)) = (x,y,p_y) \,.   
\end{equation}

\begin{figure}
  \begin{center}
      \includegraphics[width=14cm]{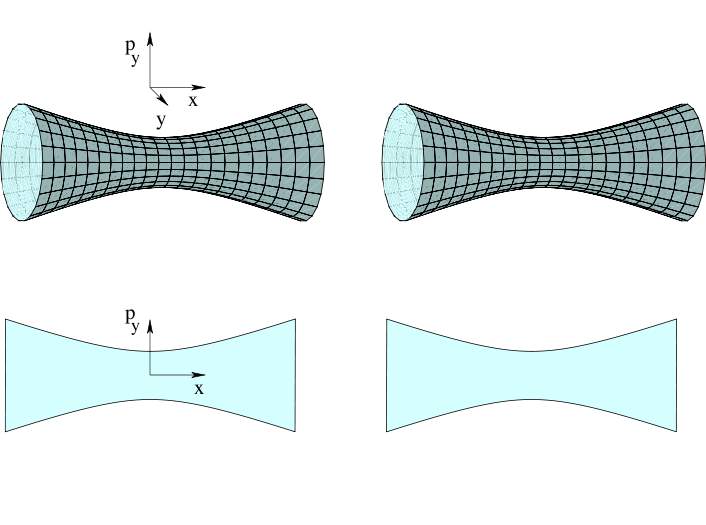} 
  \end{center}
\caption{\label{fig:esurf_standard}   Model I: the energy surface. The orientation of the $x$, $y$ and $p_y$ axes is indicated by the dreibein. (Note that the origin of the coordinate system is not the center of the dreibein.) The top left panel shows $\Sigma^{\text{I}}_+(h)$; the top left panel shows  $\Sigma^{\text{I}}_-(h)$. The two panels at the bottom show the intersections of $\Sigma^{\text{I}}_+(h)$ and  $\Sigma^{\text{I}}_-(h)$ with the plane $y=0$. 
}
\end{figure}

\noindent
The map $F^{\text{I}}$ is singular along the cylinder given by the union of the circles $ \sphere^1_{0,x}(h)$ in \eqref{eq:defequator} 
with $x\in \R$. This singular cylinder  is mapped to the quadric

\begin{equation} \label{eq:boundary_esurf_standard}
   \frac{p_y^2}{2h/\omega}  - \frac{x^2}{2h/\lambda}  + \frac{y^2}{2h/\omega} = 1
\end{equation}

\noindent
which is a single sheeted hyperboloid that is rotationally  symmetric about the $x$ axis. 
\footnote{We note that for $h<0$ the quadrics \eqref{eq:boundary_esurf_standard} are two-sheeted  hyperboloids. For $h=0$, the quadric is a `diabolo'.}
Each point in  $(x,y,p_y)\in \R^3$ with

\begin{equation}
\frac{p_y^2}{2h/\omega}  - \frac{x^2}{2h/\lambda}  + \frac{y^2}{2h/\omega} < 1
\end{equation}

\noindent
has two preimages that differ by the sign of $p_x$. We can thus represent the energy surface 
$\Sigma(h)$ in the three dimensional space, $\R^3$,  in terms of the manifolds

\begin{equation} \label{eq:esurf_model1}
\begin{split}
\Sigma^{\text{I}}_{\pm}(h) & = \{  (x,y,p_y) \in \R^3   \,:\,  x\in \R, \, (y,p_y) \in \tilde{B}^2_{\pm,x}(h) \}  \\
&= \{  (x,y,p_y) \in \R^3   \,:\,   \frac{p_y^2}{2h/\omega}  - \frac{x^2}{2h/\lambda}  + \frac{y^2}{2h/\omega} < 1 \}
  \,,
\end{split}
\end{equation}

\noindent
where   $p_x = (2h + x^2 - p_y^2 - y^2)^{1/2} $ is positive 
on $\Sigma^{\text{I}}_{+}(h)$ 
and $p_x = -(2h + x^2 - p_y^2 - y^2)^{1/2} $ is negative 
on $\Sigma^{\text{I}}_{-}(h)$.  The full energy surface $\Sigma(h)$ is obtained by gluing $\Sigma^{\text{I}}_{\pm}(h) $ along their boundaries \eqref{eq:boundary_esurf_standard} where $p_x=0$.
We show illustrations of $\Sigma^{\text{I}}_{+}(h)$ and $\Sigma^{\text{I}}_{-}(h)$ 
in the top panels of Fig.~\ref{fig:esurf_standard}.  As a consequence of the symmetry \eqref{eq:S1_symmetry} 
the components $\Sigma^{\text{I}}_{\pm}(h)$ are both rotationally symmetric about the $x$-axis. It is therefore useful to consider also the intersections of  $\Sigma^{\text{I}}_{+}(h)$ and  $\Sigma^{\text{I}}_{+}(h)$ (and of the other phase space structures discussed in the following) with the plane $y=0$ as shown
in the bottom panels of Fig.~\ref{fig:esurf_standard}.
It is easy to see the wide-narrow-wide geometry of the bottleneck type structure of the energy surface in this representation of the energy surface. The reactants region is on the left, the product regions is on the right.

\begin{figure}
  \begin{center}
      \includegraphics[width=14cm]{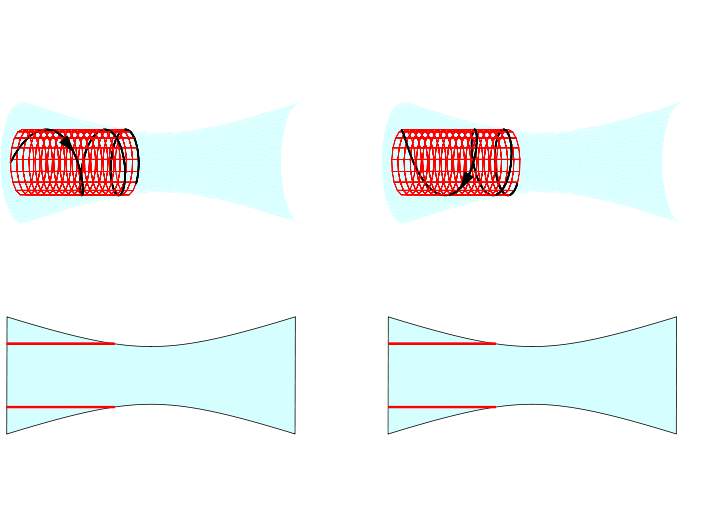} 
  \end{center}
\caption{\label{fig:non_reactive_reactants_standard}  Model I: nonreactive trajectories that stay in the reactants component. The red wireframes in the top panels indicate the two pieces of a Lagrangian cylinder  $\Lambda^-_{I,J}$ defined in \eqref{eq:def_Lambda_minus} for $I<0$ contained in $\Sigma^{\text{I}}_+(h)$ and $\Sigma^{\text{I}}_-(h)$, respectively. The black line shows one of the trajectories contained in $\Lambda^-_{I,J}$. The red lines in the bottom panels show the intersection of $\Lambda^-_{I,J}$ with the plane $y=0$.}
\end{figure}

In  Fig.~\ref{fig:non_reactive_reactants_standard}  we show the image under $F^{\text{I}}$  of a Lagrangian cylinder $\Lambda^-_{I,J}$ defined in \eqref{eq:def_Lambda_minus} for $I<0$ 
together with a nonreactive trajectory contained in this cylinder. 
Note that for a 2-\dof\ system and given values for the integral $I$ and the energy $h$, the second integral $J$ is fixed by the energy equation $K(I,J)=h$ (see \eqref{eq:def_K_2dof}). 
The trajectory approaches the bottleneck region with positive $p_x$ from 
the reactants component of the energy surface within  
$\Sigma^{\text{I}}_+(h)$  
(the left panels of Fig.~\ref{fig:non_reactive_reactants_standard}), reaches the boundary \eqref{eq:boundary_esurf_standard} of $\Sigma^{\text{I}}_+(h)$, where $p_x=0$ and the trajectory `jumps'  to $\Sigma^{\text{I}}_-(h)$  (the right panels in 
Fig.~\ref{fig:non_reactive_reactants_standard}) after  which it returns back deeper into the reactants region with $p_x<0$.

\begin{figure}
  \begin{center}
      \includegraphics[width=14cm]{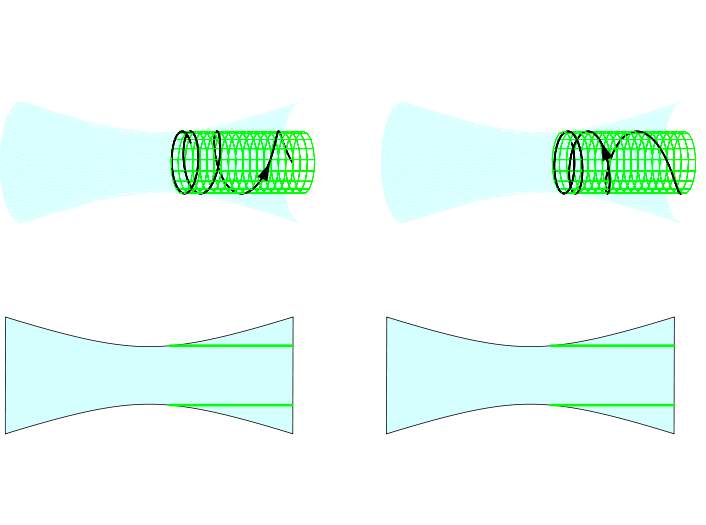} 
  \end{center}
\caption{\label{fig:non_reactive_products_standard}   Model I: nonreactive trajectories that stay in the products component.
The green wireframes in the top panels  indicate the two pieces of a Lagrangian cylinder  $\Lambda^+_{I,J}$ defined in \eqref{eq:def_Lambda_plus} for $I<0$ contained in $\Sigma^{\text{I}}_+(h)$ and $\Sigma^{\text{I}}_-(h)$, respectively. The black line shows one of the trajectories contained in $\Lambda^+_{I,J}$. The green lines in the bottom panels show the intersection of $\Lambda^+_{I,J}$ with the plane $y=0$.}
\end{figure}

Figure~\ref{fig:non_reactive_products_standard} shows the $F^{\text{I}}$  image of a Lagrangian cylinder 
$\Lambda^+_{I,J}$ for $I<0$  together with a trajectory contained in $\Lambda^+_{I,J}$. The trajectory approaches the bottleneck region  with negative $p_x$  from the products side of the energy surface within $\Sigma^{\text{I}}_-(h)$  (the right panels of Fig.~\ref{fig:non_reactive_products_standard}), reaches the boundary \eqref{eq:boundary_esurf_standard} of $\Sigma^{\text{I}}_-(h)$, where $p_x=0$ and the trajectory `jumps'  to $\Sigma^{\text{I}}_+(h)$  (the left panels in 
Fig.~\ref{fig:non_reactive_products_standard}), in which it returns back deeper in to the products region with $p_x>0$.

\begin{figure}
  \begin{center}
      \includegraphics[width=14cm]{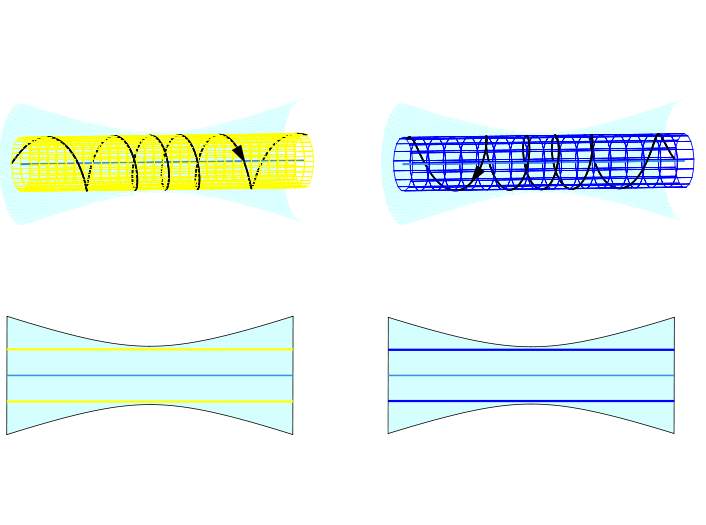} 
  \end{center}
\caption{\label{fig:reactive_cylinder_forward_backward_standard}  Model I: forward and backward reactive trajectories spiraling about the forward and backward reaction paths, respectively.
 The yellow wireframe in the top left panel indicates a Lagrangian cylinder $\Lambda^+_{I,J}$ with $I>0$ contained in $\Sigma^{\text{I}}_+(h)$; the blue wireframe in the top right panel indicates a Lagrangian cylinder $\Lambda^-_{I,J}$ with $I>0$ contained in $\Sigma^{\text{I}}_-(h)$.  The black lines show trajectories contained in the respective cylinders. 
 The blue centerlines show the forward (left) and backward (right) reaction paths.
 The bottom panels show the intersections of the Lagrangian cylinders with the plane $y=0$ together with the forward and backward reaction paths.}
\end{figure}

Figure~\ref{fig:reactive_cylinder_forward_backward_standard} 
shows the $F^{\text{I}}$ images  of Lagrangian cylinders $\Lambda^-_{I,J}$ and $\Lambda^+_{I,J}$ for $I>0$. 
For $I>0$, $\Lambda^+_{I,J}$ contains forward reactive trajectories which have $p_x>0$ and therefore are completely contained in $\Sigma^{\text{I}}_+(h)$ (the left panels in Fig.~\ref{fig:reactive_cylinder_forward_backward_standard}), and  $\Lambda^-_{I,J}$ contains backward reactive trajectories which have $p_x<0$ and therefore are completely contained in $\Sigma^{\text{I}}_-(h)$ (the right panels in Fig.~\ref{fig:reactive_cylinder_forward_backward_standard}).
The forward and backward trajectories spiral about the forward and backward reaction paths  which are also shown in Fig.~\ref{fig:reactive_cylinder_forward_backward_standard}.

\begin{figure}
  \begin{center}
      \includegraphics[width=14cm]{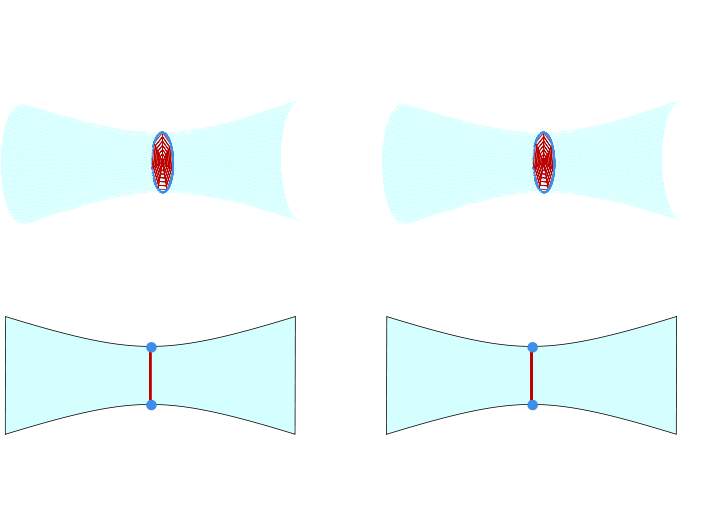} 
  \end{center}
\caption{\label{fig:DS_NHIM_standard}   Model I: the dividing surface and the NHIM.
The dark red wireframes indicate the two hemispheres  $\tsftwoDOF$  (top left panel) and $\tsbtwoDOF$ (top right panel) the dividing surface $\tstwoDOF$.  The light blue circle in the top panels shows the periodic orbit $\po$. The lower panels show the intersections of  
$\tsftwoDOF$, $\tsbtwoDOF$ and $\po$ with the plane $y=0$.
}
\end{figure}

The  images of the dividing surface,  $\tstwoDOF$, and the NHIM,  $\po$, under $F^{\text{I}}$ are shown in Fig.~\ref{fig:DS_NHIM_standard}. 
The forward hemisphere $\tsftwoDOF$ is contained in $\Sigma^{\text{I}}_+(h)$ (the left panels in Fig.~\ref{fig:DS_NHIM_standard});
the backward hemisphere $\tsbtwoDOF$ is contained in $\Sigma^{\text{I}}_-(h)$ (the right panels in Fig.~\ref{fig:DS_NHIM_standard}).
The periodic orbit $\po$ which forms the equator of $\tstwoDOF$  has $p_x=0$ and thus runs along the boundaries of $\Sigma^{\text{I}}_+(h)$ and $\Sigma^{\text{I}}_-(h)$.

\begin{figure}
  \begin{center}
      \includegraphics[width=14cm]{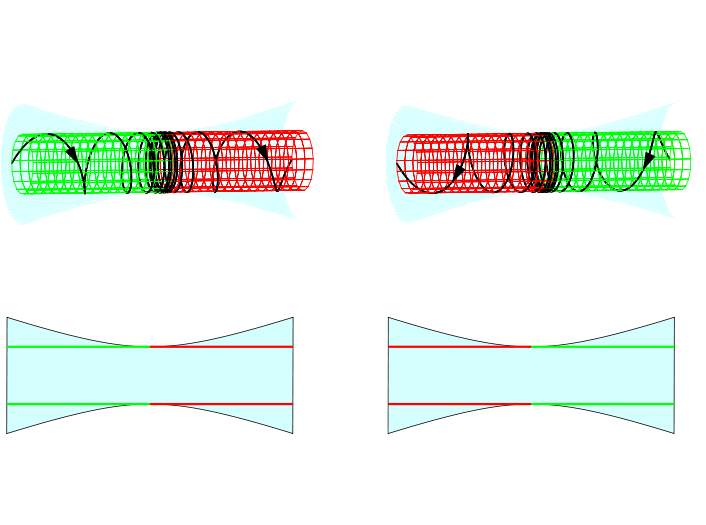} 
  \end{center}
\caption{\label{fig:forward_backward_cylinders_standard}  Model I: the stable and unstable manifolds of the NHIM. 
The wireframes in the top left panel show the branches $W^s_{\text{f}}$ (green) and $W^u_{\text{f}}$ (red), and
the wireframes in the top right panel show the branches $W^s_{\text{b}}$ (green) and $W^u_{\text{b}}$ (red) of the stable and unstable manifolds of $\po$.  For each branch, a trajectory (black line) contained in the respective branch is shown. 
The bottom panels show the intersections of the branches $W^{s/u}_{\text{f/b}}$ with the plane $y=0$.
}
\end{figure}

Figure~\ref{fig:forward_backward_cylinders_standard} shows the $F^{\text{I}}$ images of the stable and unstable manifolds  $W^{s}$ and $W^u$ of $\po$.
The forward branches $W^{s}_{\text{f}}$ and $W^{u}_{\text{f}}$ are contained in 
$\Sigma^{\text{I}}_+(h)$ (left panels in Fig.~\ref{fig:forward_backward_cylinders_standard}), and the backward branches 
$W^{s}_{\text{b}}$ and $W^{u}_{\text{b}}$ are contained in $\Sigma^{\text{I}}_-(h)$ (left panels in 
Fig.~\ref{fig:forward_backward_cylinders_standard}).  Figure~\ref{fig:forward_backward_cylinders_standard} also shows how trajectories contained in the branches $W^{s/u}_{\text{f/b}}$ asymptotically approach the periodic orbit $\po$ in the respective time 
limit $t \to \pm \infty$.
Regarding Fig.~\ref{fig:forward_backward_cylinders_standard} together with 
Fig.~\ref{fig:reactive_cylinder_forward_backward_standard} we see that
the forward reactive cylinder $W^s_{\text{f}}\cup W^u_{\text{f}}$ contained in $\Sigma^{\text{I}}_+(h)$ encloses all forward reactive trajectories, 
and the backward reactive cylinder  $W^s_{\text{b}}\cup W^u_{\text{b}}$ encloses all the backward reactive trajectories. 
Note that the $F^{\text{I}}$ images of the forward and backward reactive cylinders  are smooth along their junctions given by the periodic orbit $\po$ while in the original phase space $W^s_{\text{f}}\cup W^u_{\text{f}}$ and   $W^s_{\text{b}}\cup W^u_{\text{b}}$  have  a kink along $\po$.  This is an artifact  of the map $F^{\text{I}}$ which, as mentioned above, is singular along the cylinder $\cup_{x\in\R}  \sphere^1_{0,x}(h)$ which contains $\po=\sphere^1_{0,0}(h)$.

%%%%%%%%%%%
%%%%%%%%%%%
%\input{model_II}
%%%%%%%%%%%
%%%%%%%%%%%

\subsection{Model II:  stereographic projection of 2-spheres}
\label{sec:stereo}

\begin{figure}[!htb]
\begin{center}
  \includegraphics[width=12cm]{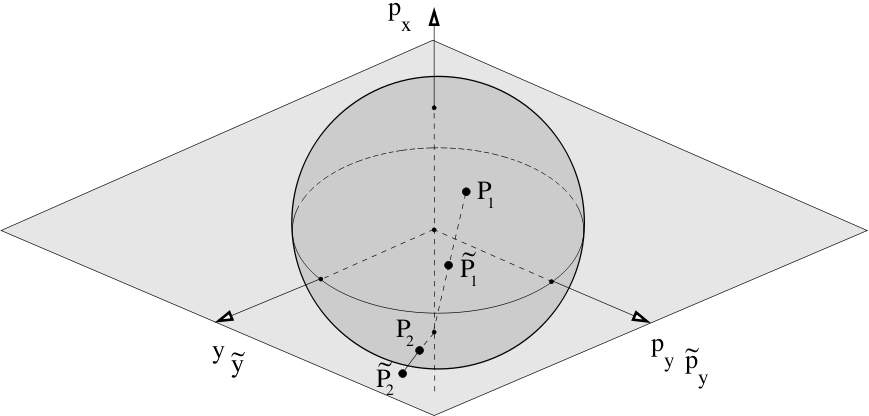}     
\end{center}
   \caption[]{\label{fig:projection_stereogr} Model II: stereographic projection of $\sphere^2_{x}(h)$ to its equatorial plane with coordinates $(\tilde{y},\tilde{p}_y)$. The points  $\tilde{P}_1$ and  $\tilde{P}_2$ are the images of the points  $P_1$ and $P_2$ under the projection \eqref{eq:def_stereo_proj}.}
\end{figure}

The main idea of the second model (Model II) is to represent the 2-spheres  $\sphere^2_{x}(h)$ defined by Eq.~\eqref{eq:def_sphere_x_h} in terms of  stereographic type projections to a plane. Recall that $\sphere^2_{x}(h)$ is given by the axisymmetric ellipsoid \eqref{eq:def_sphere_x_h}. We define a stereographic projection of this ellipsoid by projecting each point on it along the line through this point and  the South pole $(p_x, p_y, y)=(-\sqrt{(2h+\lambda x^2)/\lambda},0,0)$ of the ellipsoid 
to the equatorial plane of the ellipsoid. This projection is illustrated in Fig.~\ref{fig:projection_stereogr}.
Points on the hemisphere $B^2_{+,x}(h) $  of $\sphere^2_{x}(h)$, which we defined in \eqref{eq:defB2plus}, are projected to the circular disk bounded by the equator \eqref{eq:defequator} (e.g., the point $P_1$ in Fig.~\ref{fig:projection_stereogr}). 
The North pole is mapped to the center of this disk.
Points on the hemisphere  $B^2_{-,x}(h) $ defined in \eqref{eq:defB2minus} are projected to points in the plane outside of the disk (see, e.g., the point $P_2$ in Fig.~\ref{fig:projection_stereogr}). The South pole $(p_x, p_y, y)=(-\sqrt{(2h+\lambda x^2)/\lambda},0,0)$ is mapped to infinity. The equator $ \sphere^1_{0,x}(h) $ of   $\sphere^2_{x}(h)$ (see  \eqref{eq:defequator}) is invariant under the projection.  
Formally this  projection  is  given by
\begin{equation}\label{eq:def_stereo_proj}
\pi_{\text{sg}\,x}:\sphere^2_{x}(h)\to \overline{\R}^2\,,\quad (p_x,p_y,y) \mapsto (\tilde{y},\tilde{p}_y)\,,
\end{equation}
\noindent
where 

\begin{equation} \label{eq:stereo}
\tilde{y} = \frac{y\sqrt{(2h+\lambda x^2)/\lambda}}{\sqrt{(2h+\lambda x^2)/\lambda}+p_x}\,,\quad
\tilde{p}_y =  \frac{p_y\sqrt{(2h+\lambda x^2)/\lambda}}{\sqrt{(2h+\lambda x^2)/\lambda}+p_x}\,,
\end{equation}

\noindent
and $\overline{\R}^2 $ denotes the compactified plane $\R^2\cup \{\infty\}$.  
Note that the familiar standard stereographic projection is from a 2-sphere of unit radius to its equatorial plane where the `base point' of the projection is chosen to be the North pole rather than the South pole like in our choice. This explains the scaling factors and signs in  \eqref{eq:stereo} which differ from the standard formulae. 
We use the projection $\pi_{\text{sg}\,x}$ in \eqref{eq:def_stereo_proj} to define a map of points 
$(x,y,p_x,p_y)$ on the energy surface $\Sigma(h)$ according to 

\begin{equation}
F^{\text{II}} :  (x,y,p_x,p_y) \mapsto (x,\pi_{{\text{sg}} \, x}(p_x, p_y, y)) = (x,\tilde{y},\tilde{p}_y) \,.
\end{equation}

\noindent
The map $F^{\text{II}}$ is singular along the North poles of the spheres $\sphere^2_{x}(h)$, $x\in \R$, where $y$ and $p_y$ are zero, and $p_x$ reaches its minimal value for fixed $x$ and $h$. The set of singular points thus is the backward reaction path which is mapped to infinity. As a result the image of  $\Sigma(h)$ under 
$F^{\text{II}}$ is the Cartesian product $\R\times\overline{\R}^2$. Note that the map $F^{\text{II}}$ preserves the symmetry \eqref{eq:S1_symmetry}, or more precisely, like in Model I, manifolds which are invariant under the $\sphere^1$ symmetry action \eqref{eq:S1_symmetry} are rotationally symmetric about the $x$ axis of the image of  $F^{\text{II}}$.

\begin{figure}
  \begin{center}
      \includegraphics[width=12cm]{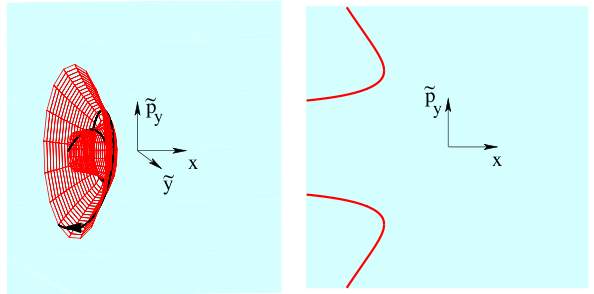} 
  \end{center}
\caption{\label{fig:nonreactive_cylinder_reactants_stereo}   Model II: nonreactive trajectories that stay in the reactants component. 
The orientation of the $x$, $\tilde{y}$ and $\tilde{p}_y$ axes is indicated by the dreibein.  The wireframe in the left panel shows a Lagrangian cylinder  $\Lambda^-_{I,J}$ defined in \eqref{eq:def_Lambda_minus} for $I<0$.
The black line shows one of the trajectories contained in $\Lambda^-_{I,J}$. The red lines in the right panel show the intersection of $\Lambda^-_{I,J}$ with the plane $\tilde{y}=0$. 
}
\end{figure}

\begin{figure}
  \begin{center}
      \includegraphics[width=12cm]{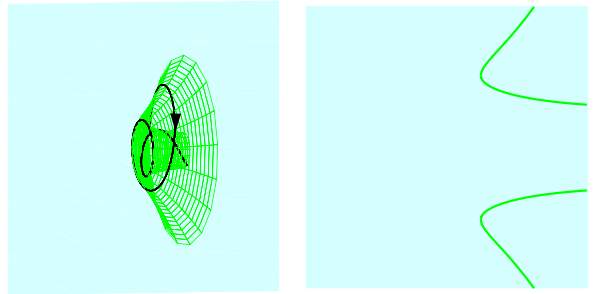} 
  \end{center}
\caption{\label{fig:nonreactive_cylinder_products_stereo}   Model II: nonreactive trajectories that stay in the products component.
The green wireframe in the left panel shows a Lagrangian cylinder $\Lambda^+_{I,J}$ 
defined in \eqref{eq:def_Lambda_plus} for $I<0$. The black line shows one of the trajectories contained in $\Lambda^+_{I,J}$.
The green lines in the right panel show the intersection of $\Lambda^+_{I,J}$ with the plane $\tilde{y}=0$. }
\end{figure}

In the following we use this representation to show geometrical visualizations of the various phase space structures 
that govern reaction dynamics analogously to Sec.~\ref{sec:model-hw}. Since the energy surface maps to 
$\R \times \overline{\R}^2$  under $F^{\text{II}}$  we omit a picture of the image of the energy surface. 
In Fig.~\ref{fig:nonreactive_cylinder_reactants_stereo}
we show the $F^{\text{II}}$  image of a Lagrangian manifold $\Lambda^-_{I,J}$ with $I<0$ together with a nonreactive trajectory contained in this cylinder. 
The trajectory comes in from the reactants component ($x\ll -1$) on the left  with positive $p_x$, spiraling  about the $x$ axis. The radius of the rotation increases as $x$ increases. After reaching a maximal $x$ value (where $p_x=0$) the trajectory returns deeper and deeper into the reactants region, still spiraling about the $x$ axis with the radius of rotation increasing as $x$ decreases. 
In Fig.~\ref{fig:nonreactive_cylinder_products_stereo} we show the analogous picture for the $F^{\text{II}}$ image of a Lagrangian manifold $\Lambda^+_{I,J}$ with $I<0$.  The trajectories come in from the products component on the right ($x\gg 1$) with negative $p_x$ undergoing rotations with increasing radius about the $x$ axis as $x$ decreases. After reaching a minimal $x$ value, the trajectory returns deeper and deeper into the products region, still undergoing rotation about the $x$ axis, where the radius of the rotation now increases as $x$ increases.

\begin{figure}
  \begin{center}
      \includegraphics[width=12cm]{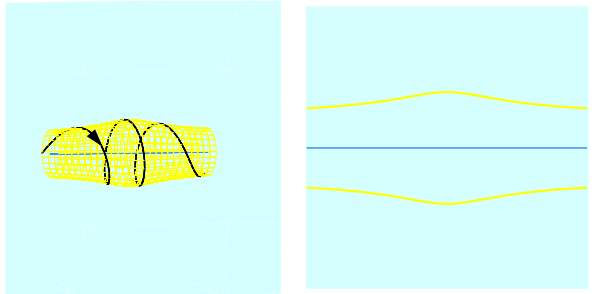} 
  \end{center}
\caption{\label{fig:reactive_cylinder_forward_stereo}   Model II: forward reactive trajectories spiraling about the forward reaction path.
 The yellow wireframe in the  left panel indicates a Lagrangian cylinder $\Lambda^+_{I,J}$ with $I>0$.  The black line shows a trajectory contained in  $\Lambda^+_{I,J}$. The blue line shows the forward reaction path.
 The right panel shows the intersection of $\Lambda^+_{I,J}$  with the plane $\tilde{y}=0$ together with the forward reaction path.}
\end{figure}

Figure~\ref{fig:reactive_cylinder_forward_stereo} shows  the $F^{\text{II}}$ image of a Lagrangian cylinder  $\Lambda^+_{I,J}$ with $I>0$ together with the forward reaction path which coincides with the $x$ axis.  
Trajectories contained in $\Lambda^+_{I,J}$ evolve from the reactants to the products spiraling about the forward reaction path.
\begin{figure}
  \begin{center}
      \includegraphics[width=12cm]{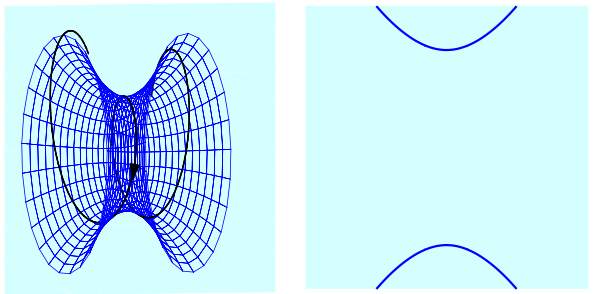} 
  \end{center}
\caption{\label{fig:reactive_cylinder_backward_stereo}   Model II: backward reactive trajectories.
 The blue wireframe in the  left panel indicates a Lagrangian cylinder $\Lambda^-_{I,J}$ with $I>0$.  The black line shows a trajectory contained in  $\Lambda^-_{I,J}$.
 The right panel shows the intersection of $\Lambda^-_{I,J}$  with the plane $\tilde{y}=0$.}
\end{figure}
Figure~\ref{fig:reactive_cylinder_backward_stereo} shows the $F^{\text{II}}$ image of a Lagrangian cylinder  $\Lambda^-_{I,J}$ with $I>0$ that contains backward reactive trajectories. These trajectories spiral with much larger radii of rotation than the forward reactive trajectories in Fig.~\ref{fig:reactive_cylinder_forward_stereo}. In fact, since the backward reaction path is mapped to the line $\R \cup \infty$
 the backward reactive trajectories can again be considered to spiral about the backward reaction path.

\begin{figure}
  \begin{center}
      \includegraphics[width=12cm]{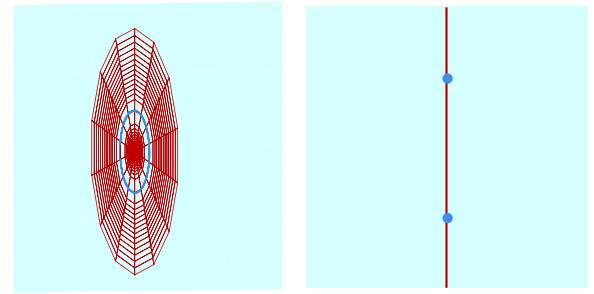} 
  \end{center}
\caption{\label{fig:DS_NHIM_stereo}  Model II: dividing surface and NHIM. The dark red wireframe in the left panel shows the dividing surface $\tstwoDOF$. The periodic orbit $\po$ (blue circle) divides it into the forward hemisphere $\tsftwoDOF$ (inner circular disk) and the backward hemisphere  $\tsbtwoDOF$ (outer infinite annulus). The right panel shows the intersection of $\tstwoDOF$ and $\po$ with the plane $\tilde{y}=0$.
}
\end{figure}

\begin{figure}
  \begin{center}
      \includegraphics[width=12cm]{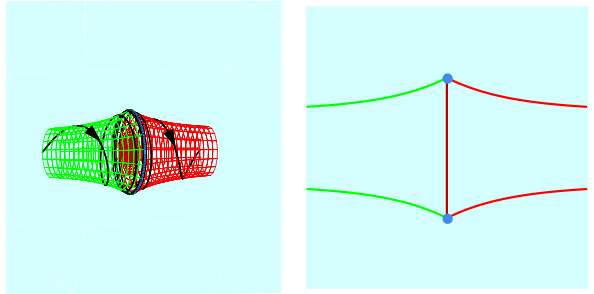} 
  \end{center}
\caption{\label{fig:forward_cylinder_stereo}  Model II: forward reactive cylinder, forward dividing surface and NHIM. The green and red wireframes in the left panel indicate the branches $W^s_{\text{f}} $ and $W^u_{\text{f}} $ of the stable resp. unstable manifolds of the NHIM $\po$ (blue circle). The black lines show trajectories contained in $W^s_{\text{f}}  $ and $W^u_{\text{f}} $. The dark red wireframe at the center which is difficult to see shows the forward hemisphere $\tsftwoDOF$ of the dividing surface. The right panel shows the intersections of the structures on the right with the plane $\tilde{y}=0$.
}
\end{figure}

\begin{figure}
  \begin{center}
      \includegraphics[width=12cm]{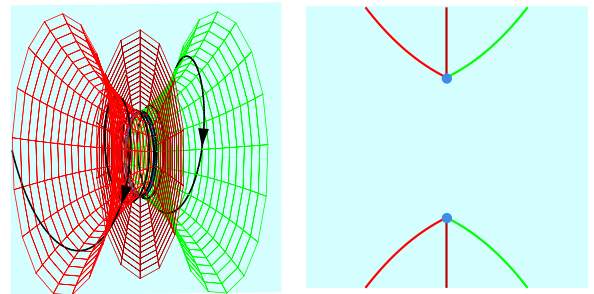} 
  \end{center}
\caption{\label{fig:backward_cylinder_stereo}   Model II: backward reactive cylinder, backward dividing surface and NHIM. The green and red wireframes in the left panel indicate the branches $W^s_{\text{b}} $ and $W^u_{\text{b}} $ of the stable resp. unstable manifolds of the NHIM $\po$ (blue circle). The black lines show trajectories contained in $W^s_{\text{b}}  $ and $W^u_{\text{b}} $. The dark red wireframe   shows the backward hemisphere $\tsbtwoDOF$ of the dividing surface. The right panel shows the intersections of the structures on the right with the plane $\tilde{y}=0$.
}
\end{figure}

In Fig.~\ref{fig:DS_NHIM_stereo} we show the images of the dividing surface $\tstwoDOF$ and the NHIM  $\po$ under $F^{\text{II}}$. 
The dividing surface $\tstwoDOF$ maps to  the compactified plane $x=0$. The $F^{\text{II}}$ image of the NHIM, $\po$, divides it into the forward 
hemisphere $\tsftwoDOF$ which appears as a circular disk in the plane $x=0$ centered about the origin and the backward hemisphere $\tsbtwoDOF$ which appears as an annulus that extends to the boundary of $\R^2$ at infinity.   
The $F^{\text{II}}$ image of the forward reactive trajectories in Fig.~\ref{fig:reactive_cylinder_forward_stereo}
which are located near the $x$ axis cross the $F^{\text{II}}$ image of $\tsftwoDOF$, and the $F^{\text{II}}$ image of the backward reactive trajectories in Fig.~\ref{fig:reactive_cylinder_backward_stereo} which stay away from the $x$ axis cross the $F^{\text{II}}$ image of $\tsbtwoDOF$.
The two types of trajectories are enclosed by the forward and backward reactive cylinders  $W^s_{\text{f}}\cup W^u_{\text{f}}$ and  $W^s_{\text{b}}\cup W^u_{\text{b}}$ whose $F^{\text{II}}$ images are shown in Figs.~\ref{fig:forward_cylinder_stereo} and \ref{fig:backward_cylinder_stereo}, respectively.

Note that in this geometrical representation of the energy surface (Model II) the bottleneck area of  $\Sigma(h)$ associated with the  reaction region containing the dividing surface is not so pronounced as in Model I. 

%%%%%%%%%
%%%%%%%%%
%\input{model_III}
%%%%%%%%%
%%%%%%%%%

\subsection{Model III: McGehee representation}
\label{sec:McGehee}

The main idea of the third model (Model III) of the energy surface $\Sigma(h)$ is to represent the family of  (topological)  2-spheres, $\sphere^2_{x}(h)$, in 
Eq.~\eqref{eq:def_sphere_x_h} parametrized by $x\in\R$ as a `nested'  set of concentric 2-spheres in $\R^3$. This can be achieved by projecting each of the 2-spheres, $\sphere^2_{x}(h)$ to a 2-sphere of a radius which depends on $x$. One way to realize this is to map a point $P=(p_x,p_y, y)$ on  the $\sphere^2_{x}(h)$ 
according to

\begin{equation} \label{eq:def_pi_McG}
  \pi_{\text{McG}\,x}: (y,p_x,p_y) \mapsto  (y, p_x, p_y) \frac{s(x)}{ \norm{ (y, p_x, p_y)  } } =: (\hat{y},\hat{p}_x,\hat{p}_y) \,.
\end{equation}

\begin{figure}[!htb]
\begin{center}
  \includegraphics[width=8cm]{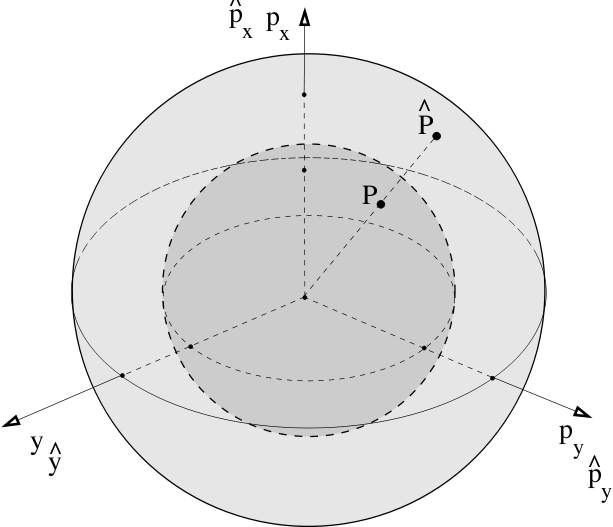}     
\end{center}
   \caption[]{\label{fig:projection_McGehee}  Model III: McGehee projection of a point $P=(y,p_x,p_y)$ on the (topological) 2-sphere $\sphere^2_{x}(h)$ (inner dark shaded sphere) to the point $\hat{P}=P=(\hat{y},\hat{p}_x,\hat{p}_y)$ on the sphere of radius $s(x)$  (outer light shaded sphere) with $s(x)$ defined in \eqref{eq:def_s}. Note that, depending on $x$, the outer sphere could also be inside of the inner sphere, or the two could overlap.}
\end{figure}

\noindent
Here $\norm{\cdot }$ denotes the Euclidean norm, i.e. $ \norm{ (p_x, p_y, y)} = (p_x^2+p_y^2+ y^2)^{1/2}$, and $s$ is the scaling function

\begin{equation} \label{eq:def_s}
s(x) :=  1+\frac{1}{2}  \tanh x
\end{equation}

\noindent
which maps the domain $x\in \R$ monotonicly  to the interval $(1/2,3/2)$. 
The projection $\pi_{\text{McG}\,x}$ is illustrated in Fig.~\ref{fig:projection_McGehee}. We name this projection after  McGehee 
who first used this type of construction to present an energy surface of the structure $\sphere^2\times \R$ \cite{McGehee69}. 
A similar construction can also be found in the work by MacKay \cite{MacKay1}.

We use  $ \pi_{\text{McG}\,x}$ in \eqref{eq:def_pi_McG} to define a map $F^{\text{III}}$ 
of points  $(x,y,p_x,p_y)$ on the energy surface $\Sigma(h)$ to $\R^3$ according to 
\begin{equation}
F^{\text{III}} :  (x,y,p_x,p_y) \mapsto  \pi_{\text{McG} \, x}(p_x,p_y, y)=:(\hat{y},\hat{p}_x,\hat{p}_y)\,.
\end{equation}
Like the maps $F^{\text{I}}$ and $F^{\text{II}}$, the map $F^{\text{III}}$ also preserves the $\sphere^1$ symmetry. In Model III, manifolds that are invariant under the symmetry action \eqref{eq:S1_symmetry} have $F^{\text{III}}$ images that are symmetric about the $\hat{p}_x$ axis.

\begin{figure}
  \begin{center}
      \includegraphics[width=12cm]{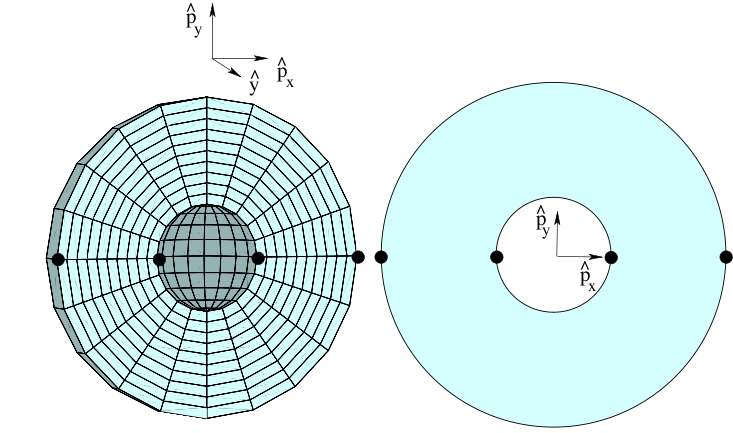} 
  \end{center}
\caption{\label{fig:esurf_McGehee}   Model III: the energy surface. The left panel shows the energy surface in the threedimensional space with coordinates $(\hat{y},\hat{p}_y,\hat{p}_x)$. The orientation of the  $\hat{y}$, $\hat{p}_y$ and $\hat{p}_x$ axes are indicated by the dreibein. (Note that the center of the coordinate system is not the center of the dreibein.) The black points mark the points  
$(\hat{y},\hat{p}_y,\hat{p}_x)=(0,0,\pm1/2)$  and  $(\hat{y},\hat{p}_y,\hat{p}_x) = (0,0,\pm 3/2)$ (see text). The right panel shows the intersection of the energy surface with the plane $\hat{y}=0$.}
\end{figure}

The image of the energy surface  $\Sigma(h)$ of  topology $\R\times \sphere^2$ under the map $F^{\text{III}}$ is the `spherical shell' consisting of the open ball  of radius $3/2$ in $\R^3$  minus the closed ball of radius $1/2$. This image is shown in Fig.~\ref{fig:esurf_McGehee}.
The reactants region $x\ll -1$ is contained in the outer region of this spherical shell; 
the products region $x\gg 1$ is contained in the inner region.
\begin{figure}
  \begin{center}
      \includegraphics[width=12cm]{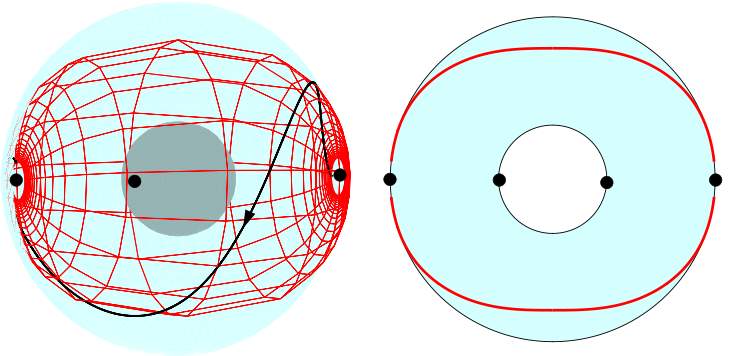} 
  \end{center}
\caption{\label{fig:nonreactive_cylinder_reactants}   Modell III: nonreactive trajectories that stay in the reactants component. The red wireframe in the left panel indicates a Lagrangian cylinder  $\Lambda^-_{I,J}$ defined in \eqref{eq:def_Lambda_minus} for $I<0$.  The black line shows one of the trajectories contained in $\Lambda^-_{I,J}$. The red lines in the right panel show the intersection of $\Lambda^-_{I,J}$ with the plane $\hat{y}=0$.}
\end{figure}
In Fig.~\ref{fig:esurf_McGehee} we also marked the points    $(\hat{y},\hat{p}_y,\hat{p}_x)=(0,0,\pm1/2)$ and  $(\hat{y},\hat{p}_y,\hat{p}_x) = (0,0,\pm 3/2)$  to which the $F^{\text{III}}$ images of all trajectories  converge as time goes to $-\infty$ or $+\infty$ (as long as the trajectory is unbounded in the respective limit). This is a consequence of the conservation of the integrals $I$ and $J$ (the energy in the $x$ and $y$ degrees of freedom, respectively) which implies that $y$ and $p_y$ are bounded for each trajectory whereas   $p_x$ diverges as $x$ goes to $-\infty$ or $+\infty$. In Fig.~\ref{fig:nonreactive_cylinder_reactants} we show the example of the $F^{\text{III}}$ image of a Lagrangian cylinder  $\Lambda^-_{I,J}$ defined for $I<0$. The trajectories in   $\Lambda^-_{I,J}$ come in from deep in the reactants region where 
 $x\ll -1$ and $p_x\gg1$ which is asymptotically (in the limit $t\to -\infty$) mapped to $(\hat{y},\hat{p}_y,\hat{p}_x)=(0,0,3/2)$ under $F^{\text{III}}$ and return deep into the reactants region where $x\ll-1$ and $p_x\ll -1$ which is asymptotically (in the limit $t\to \infty$) mapped to $(\hat{y},\hat{p}_y,\hat{p}_x)=(0,0,-3/2)$ under $F^{\text{III}}$.
 
\begin{figure}
  \begin{center}
      \includegraphics[width=12cm]{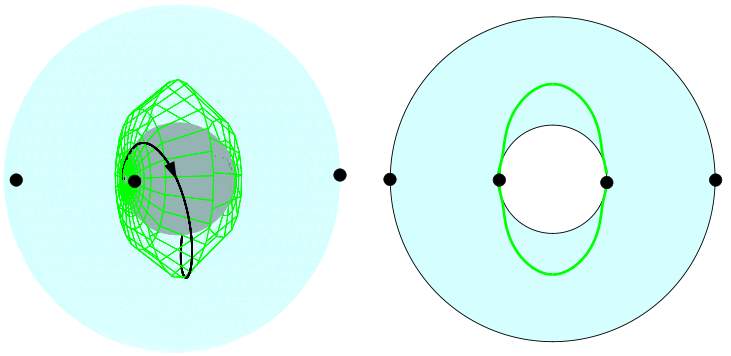} 
  \end{center}
\caption{\label{fig:nonreactive_cylinder_products_McGehee}    Modell III: nonreactive trajectories that stay in the products component. The green wireframe in the left panel indicates a Lagrangian cylinder  $\Lambda^+_{I,J}$ defined in \eqref{eq:def_Lambda_plus} for $I<0$.  The black line shows one of the trajectories contained in $\Lambda^+_{I,J}$. The green lines in the right panel show the intersection of $\Lambda^+_{I,J}$ with the plane $\hat{y}=0$.}
\end{figure}

In Fig.~\ref{fig:nonreactive_cylinder_products_McGehee} we show the $F^{\text{III}}$ image of a Lagrangian cylinder $\Lambda^+_{I,J}$ with  $I<0$ which contains nonreactive trajectories in the products region.  They come in from deep in the products region  $x\gg1$ with $p_x\ll-1$ which under   $F^{\text{III}}$ is mapped to $(\hat{y},\hat{p}_y,\hat{p}_x)=(0,0,-1/2)$ and move back deep into the products region  $x\gg1$ with $p_x\gg1$ which under $F^{\text{III}}$ is mapped to $(\hat{y},\hat{p}_y,\hat{p}_x)=(0,0,1/2)$.

\begin{figure}
  \begin{center}
      \includegraphics[width=12cm]{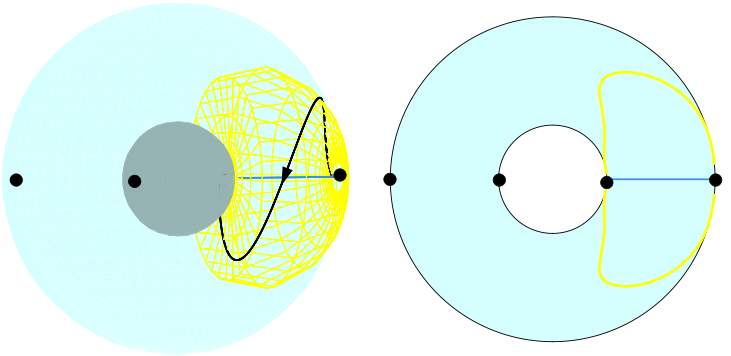} 
  \end{center}
\caption{\label{fig:reactive_cylinder_forward_McGehee}   Model III:  forward reactive trajectories spiraling about the forward reaction path.  The yellow wireframe in the  left panel indicates a Lagrangian cylinder $\Lambda^+_{I,J}$ with $I>0$.  The black line shows a trajectory contained in  $\Lambda^+_{I,J}$. The blue line shows the forward reaction path.
 The right panel shows the intersection of $\Lambda^+_{I,J}$  with the plane $\hat{y}=0$ together with the forward reaction path.}
\end{figure}

Fig.~\ref{fig:reactive_cylinder_forward_McGehee}  shows the $F^{\text{III}}$ image of a Lagrangian cylinder $\Lambda^+_{I,J}$ which has $I>0$.  The Lagrangian manifold  consists of forward reactive trajectories which spiral about the forward reaction path that is also shown in Fig.~\ref{fig:reactive_cylinder_forward_McGehee}.
\begin{figure}
  \begin{center}
      \includegraphics[width=12cm]{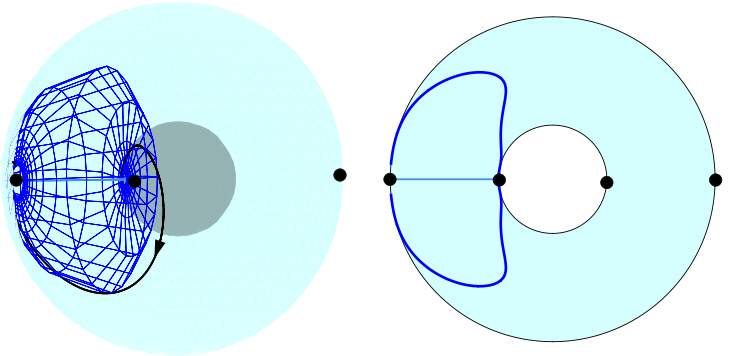} 
  \end{center}
\caption{\label{fig:reactive_cylinder_backward_McGehee}  Model III: backward reactive trajectories spiraling about the backward reaction path.
 The blue wireframe in the  left panel indicates a Lagrangian cylinder $\Lambda^-_{I,J}$ with $I>0$.  The black line shows a trajectory contained in  $\Lambda^-_{I,J}$ spiraling about the backward reaction path marked in blue.
 The right panel shows the intersection of $\Lambda^-_{I,J}$  with the plane $\hat{y}=0$ together with the backward reaction path.}
\end{figure}
\begin{figure}
  \begin{center}
      \includegraphics[width=12cm]{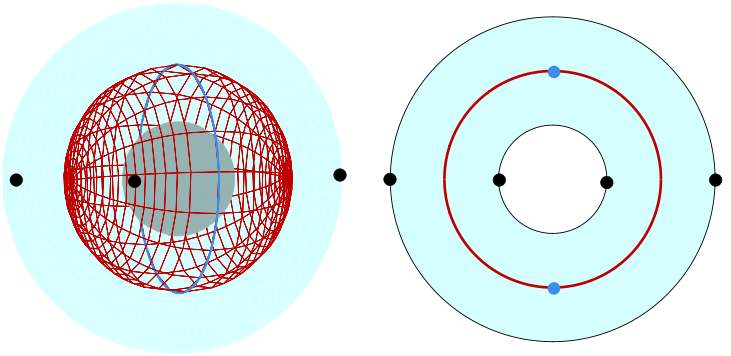} 
  \end{center}
\caption{\label{fig:DS_NHIM_McGehee}  Model III: dividing surface and NHIM. The dark red wireframe in the left panel shows the dividing surface $\tstwoDOF$. The periodic orbit $\po$ (blue circle) divides it into the forward hemisphere $\tsftwoDOF$ (the hemisphere on the right) and the backward hemisphere  $\tsbtwoDOF$ (the hemisphere on the left). The right panel shows the intersection of $\tstwoDOF$ and $\po$ with the plane $\hat{y}=0$.}
\end{figure}
Fig.~\ref{fig:reactive_cylinder_backward_McGehee} shows the analogous image of a Lagrangian cylinder $\Lambda^-_{I,J}$ which has $I>0$ and hence consists of backward reactive trajectories.

\begin{figure}
  \begin{center}
      \includegraphics[width=12cm]{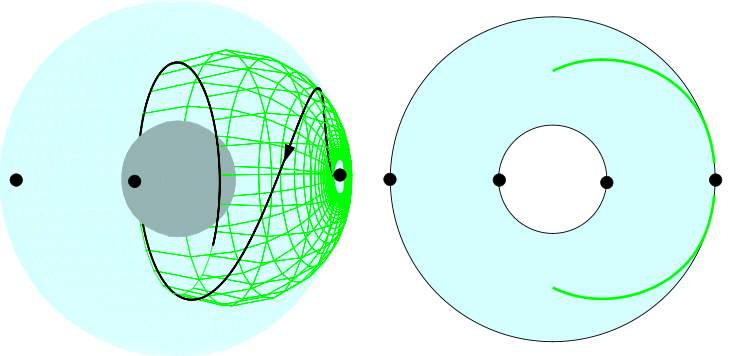} 
  \end{center}
\caption{\label{fig:Wsf_McGehee}   Model III: the forward branch of the stable manifold. 
The green wireframe in the left panel shows $W^s_{\text{f}}$. The black line shows a trajectory contained in $W^s_{\text{f}}$. The green lines in the right panels show the intersection of  $W^s_{\text{f}}$ with the plane $\hat{y}=0$.}
\end{figure}

\begin{figure}
  \begin{center}
      \includegraphics[width=12cm]{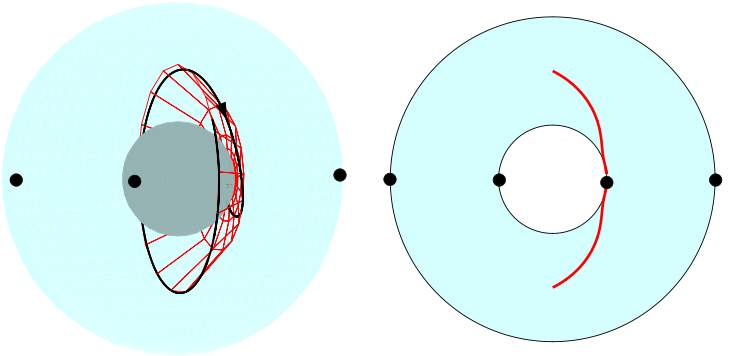} 
  \end{center}
\caption{\label{fig:Wuf_McGehee}  Model III: the forward branch of the unstable manifold. 
The red wireframe in the left panel shows $W^u_{\text{f}}$. The black line shows a trajectory contained in $W^u_{\text{f}}$. The red lines in the right panels show the intersection of  $W^u_{\text{f}}$ with the plane $\hat{y}=0$.}
\end{figure}
\begin{figure}
  \begin{center}
      \includegraphics[width=12cm]{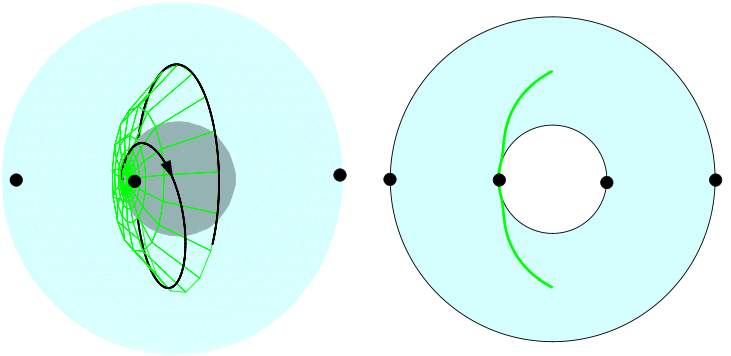} 
  \end{center}
\caption{\label{fig:Wsb_McGehee}     Model III: the backward branch of the stable manifold. 
The green wireframe in the left panel shows $W^s_{\text{b}}$. The black line shows a trajectory contained in $W^s_{\text{b}}$. The green lines in the right panels show the intersection of  $W^s_{\text{b}}$ with the plane $\hat{y}=0$.}
\end{figure}
\begin{figure}
  \begin{center}
      \includegraphics[width=12cm]{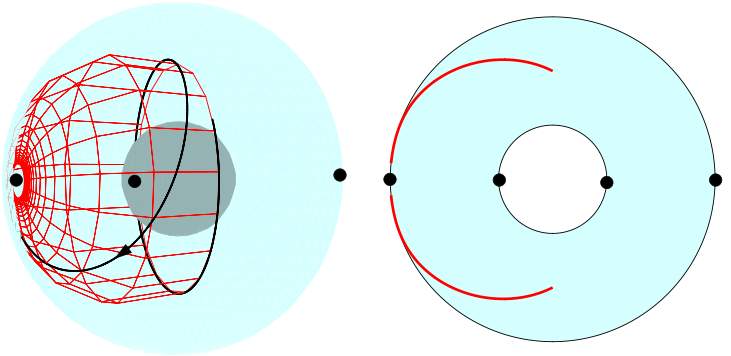} 
  \end{center}
\caption{\label{fig:Wub_McGehee}   Model III: the backward branch of the unstable manifold. 
The red wireframe in the left panel shows $W^u_{\text{b}}$. The black line shows a trajectory contained in $W^u_{\text{b}}$. The red lines in the right panels show the intersection of  $W^u_{\text{b}}$ with the plane $\hat{y}=0$.}
\end{figure}
\begin{figure}
  \begin{center}
      \includegraphics[width=12cm]{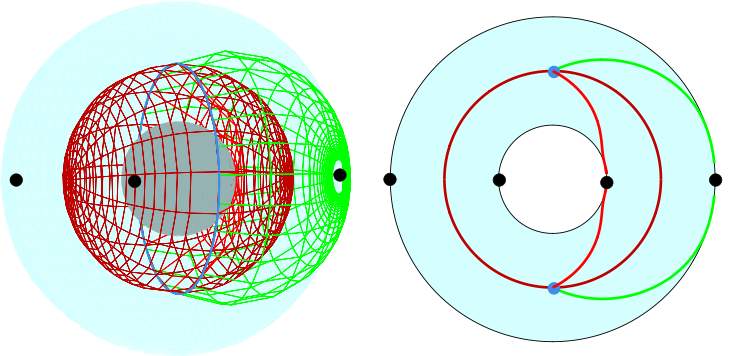} 
  \end{center}
\caption{\label{fig:forward_cylinder_McGehee}   Model III: the forward reactive cylinder, the dividing surface and the NHIM as a superposition of Figs.~\ref{fig:DS_NHIM_McGehee}, \ref{fig:Wsf_McGehee} and \ref{fig:Wuf_McGehee}.
}
\end{figure}
\begin{figure}
  \begin{center}
      \includegraphics[width=12cm]{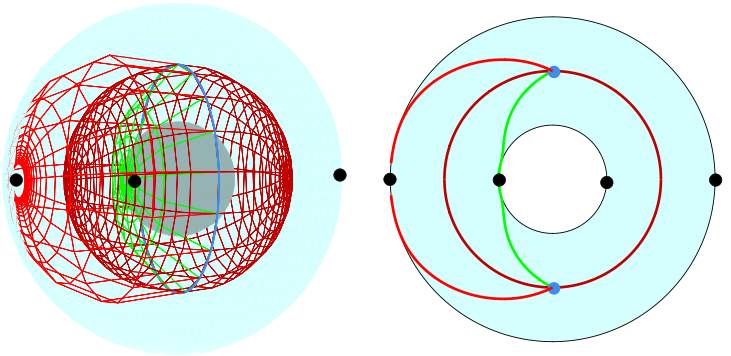} 
  \end{center}
\caption{\label{fig:backward_cylinder_McGehee}  Model III: the backward reactive cylinder, the dividing surface and the NHIM as a superposition of Figs.~\ref{fig:DS_NHIM_McGehee}, \ref{fig:Wsb_McGehee} and \ref{fig:Wub_McGehee}.}
\end{figure}

For clarity we show the $F^{\text{III}}$ images of the branches $W^s_{\text{f}}$, $W^u_{\text{f}}$,
$W^s_{\text{b}}$ and  $W^u_{\text{b}}$ of the stable and unstable manifolds of the NHIM separately 
in Figs.~\ref{fig:Wsf_McGehee}-\ref{fig:Wub_McGehee}.
The $F^{\text{III}}$ image of the union $W^s_{\text{f}} \cup W^u_{\text{f}}$ forms the forward reactive cylinder  in Model III. It encloses the $F^{\text{III}}$ image of all forward reactive trajectories as can be seen from comparing Fig.~\ref{fig:reactive_cylinder_forward_McGehee} with Fig.~\ref{fig:forward_cylinder_McGehee}. Similarly the $F^{\text{III}}$ image of the backward reactive cylinder $W^s_{\text{b}} \cup W^u_{\text{b}}$ encloses the $F^{\text{III}}$ image of all backward reactive cylinder as can be seen from comparing Fig.~\ref{fig:reactive_cylinder_backward_McGehee} with Fig.~\ref{fig:backward_cylinder_McGehee}.


%%%%%%%%%%
%%%%%%%%%%
%\input{nonlinear}
%%%%%%%%%%
%%%%%%%%%%

\subsection{Implications for Nonlinear Hamiltonian Vector Fields}
\label{sec:nonlinear_case}

We have chosen to develop the three models of the phase space structures near a saddle $\times$ centre equilibrium  by  considering a quadratic, two degree of freedom Hamiltonian which is the normal form, through second order, in the neighborhood of such an equilibrium point.  The advantage of this is that is provided very explicit formulae for the phase space structures that allowed straightforward graphical representation. 
However, precisely the same geometrical structures  exist for nonlinear Hamiltonian vector fields in the neighborhood of a saddle $\times$ center equilibrium point.  In the nonlinear case the phase space structures can again be described by explicit mathematical formulae using a normal expansion at the saddle $\times$ center equilibrium  (see Sec.~\ref{sec:structures_explicit}, and \cite{WaalkensSchubertWiggins08} for a detailed discussion). 
%
% (where the  purely imaginary eigenvalues associated with the center directions satisfy a nonresonance condition). 
% 
%Formulae for these geometric structures are given in \cite{WaalkensSchubertWiggins08}, 
%
For a given system, this in general leads a Hamilton function which is a power series rather than a linear function of the integrals  $I$ and $J$ (see \eqref{eq:def_K_2dof}). The coefficients of this powers series are obtained from 
the normal form calculations for an explicit systems. 

Nevertheless, the normal form theory allows us to conclude that the same geometrical structures, having the same interpretation and meaning, hold for more general, nonlinear Hamiltonian vector fields in the neighborhood of a saddle  center-$\ldots$-center equilibrium point, but obtaining the explicit formulae will require a normal form calculation for a given system. It is in this sense that restricting ourselves to quadratic Hamiltonians is, indeed, ''without loss of generality'', and it provides the  method and approach for visualizing the same phase space structures governing reaction dynamics in more general systems.

%%%%%%%%%%%
%%%%%%%%%%%
%\input{model_ndof}
%%%%%%%%%%%
%%%%%%%%%%%

%-----------------------------------------------------------------------
\section{A model of the $n$-DoF system in the space of the integrals of motion}
\label{sec:ndof}
%-----------------------------------------------------------------------

For a system with more than 2 degrees of freedom, the dimensionality of the phase space structures discussed in 
Sec.~\ref{sec:summary_phases_space_structures} is too high to allow for a similar explicit visualization like for  the 2-\dof models in 
Sec.~\ref{sec:2dofModels}.  Instead of projecting the phase space structures to the planes of the coordinates and conjugate momenta of the normal form like in Sec.~\ref{sec:structures_explicit} it is useful to present  
the phase space structures for $n$ \dof in the space of the integrals 
$I,J_2,\ldots,J_n$.  This can be accomplished using the momentum map ${\cal M}$ defined in \eqref{eq:def_momentum_map} in 
Sec.~\ref{sec:Lagrange_foliation}.  The discussion here follows \cite{WaalkensSchubertWiggins08}. 

\begin{figure}[htb!]
\begin{center}
\includegraphics[width=8cm]{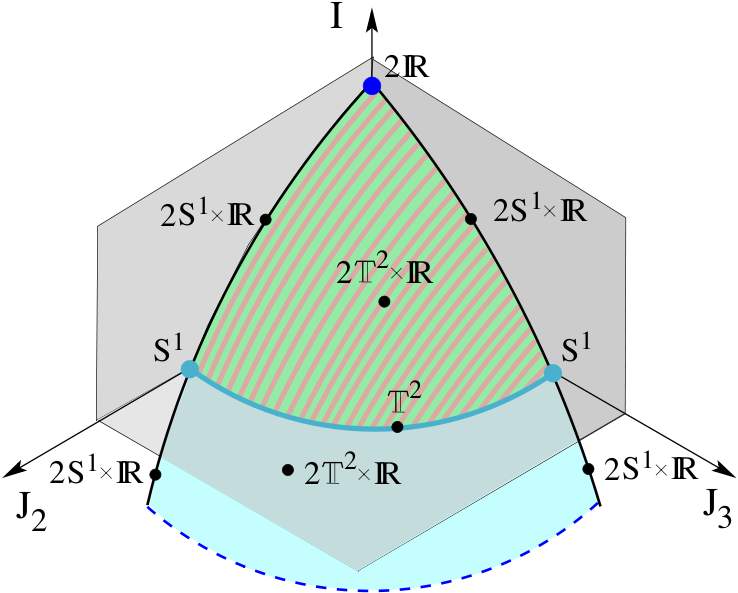}
\caption{Sketch of the image of the energy surface  of energy  $h>0$ under
the momentum map ${\cal M}$ in Equation~\eqref{eq:def_momentum_map}
in the space of the integrals $I$ and $J_k$,
$k=2,\ldots,n$, for the case of $n=3$ degrees of freedom.
The green/dark pink brindled piece of the image of the energy surface has
$I>0$; the turquoise piece has $I<0$.
The intersections with the
planes $I=0$, $J_2=0$ and $J_3=0$ (pieces of which are visualised by semitransparent planes
for clarity) form the bifurcation diagram of
the energy surface. The image of the energy surface is not bounded in the
direction of negative $I$ as indicated by the dashed line  at the bottom.
The topology of the fibres ${\cal M}^{-1}(I,J_2,J_3)$ is
indicated for the various points  $(I,J_2,J_3)$ marked by dots.
The fibre of a point  $(I,J_2,J_3)$ with $I\ne 0$ consists of two
disconnected manifolds as indicated by the factor of 2. The fibre of
a point  $(I,J_2,J_3)$ with $I = 0$ consists of a single
connected manifold.
}
\label{fig:momentummap}
\end{center}
\end{figure}

As mentioned in Sec.~\ref{sec:Lagrange_foliation}  
a fibre  (the joint level set of the integrals $I,J_2,\ldots,J_n$  in phase space) is called singular if it contains one (or more) irregular point(s).
The image of all the singular fibres under the momentum map is called the
\emph{bifurcation diagram}. It is easy to see that the bifurcation diagram
consists of the set of $(I,J_2,\ldots,J_n)$, where one or more of the integrals vanish.
In Fig.~\ref{fig:momentummap} we show the image of the energy surface with
energy $h>0$ under the momentum map ${\cal M}$  in the space of the
integrals for $n=3$ degrees of freedom.

The bifurcation diagram (of the energy
surface) consists of the intersections of the image of the energy surface (the
turquoise and green/dark ping brindled surface in Fig.~\ref{fig:momentummap})
with one of the planes $I=0$, $J_2=0$ or $J_3=0$.
Upon approaching one of the edges that have $J_2=0$ or $J_3=0$ the
circle in the plane $(q_2,p_2)$ or $(q_3,p_3)$, respectively, shrinks to a
point, and accordingly
the regular fibres $\T^{2}\times \R$ which form the Lagrange manifolds $\Lambda^+_{I,J_2,\ldots,J_n}$ and $\Lambda^-_{I,J_2,\ldots,J_n}$ (see \eqref{eq:def_Lambda_plus}~and~\eqref{eq:def_Lambda_minus}) reduce to cylinders or `tubes'
$\sphere^{1}\times \R$. At the top corner in  Fig.~\ref{fig:momentummap} both
$J_2$ and $J_3$ are zero.
Here both circles in the centre planes $(q_2,p_2)$ and $(q_3,p_3)$ have
shrinked to points.
The corresponding singular fibre consists of two
lines, $\R$, which are the forward and backward reaction paths, respectively (see
also Fig.~\ref{fig:volumes}).

The  fibres mentioned so far all have $I\ne 0$ and each consist of a pair of two
disconnected components.
For $I<0$,  one member of each pair is located on the reactants
side and the other on the products side of the dividing surface.
For $I>0$, one member of each pair consists of trajectories
evolving from reactants to products and the other member consists of
trajectories that evolve from products to reactants.
In fact the two members of a fibre which has $I>0$ are contained in the energy
surface volume enclosed by the forward and backward reactive spherical
cylinders $W_f(h)$ and $W_b(h)$, see Fig.~\ref{fig:volumes}.
For this reason we marked the piece of the image of the energy surface under
the momentum map which has $I>0$ by the same green/dark pink colour in
Fig.~\ref{fig:momentummap} that we used Fig.~\ref{fig:volumes}.
Green corresponds to forward reactive trajectories and dark pink corresponds
to backward reactive trajectories. Under the momentum map these trajectories
have the same image.

The light blue line in Fig.~\ref{fig:momentummap} which has $I=0$
is  the image of the NHIM under the momentum map. For three
degrees of freedom, the NHIM is a 3-dimensional sphere, and as
we will discuss in more detail in  Sec.~\ref{sec:nhim} and indicated in
Fig.~\ref{fig:momentummap} it is foliated by a one-parameter
family of invariant 2-tori which shrink to periodic orbits, i.e.
circles, $\sphere^1$, at the end points of the parameterisation interval.

%%%%%%%%%%%
%%%%%%%%%%%
%\input{model_nhim}
%%%%%%%%%%%
%%%%%%%%%%%

%-----------------------------------------------------------------------
\section{Structure of the NHIM for $n$-DoF}\label{sec:nhim}
%-----------------------------------------------------------------------

We have described how the NHIM, its stable and unstable
manifolds, and the dividing surface, sit within the phase space
and  govern the reaction dynamics in several 2DoF geometrical models. However, we have not given any indication of the structure of the NHIM itself, except in the $2$-DoF case, in which the NHIM is a
single periodic orbit. In this section we discuss various ways of visualizing the geometrical structure and dynamics associated with the NHIM.

Recall from Sec.~\ref{sec:structures_explicit} that in terms of the local decoupling provided by the normal form  into reaction coordinates and bath
modes, points on the NHIM, $\nhim$, contain zero `energy' in the reactive
(saddle) mode (i.e., $I=0$); all `energy' is contained in the bath (centre) modes, which results in
oscillatory, quasiperiodic motion with $n-1$ independent frequencies. 
As also mentioned in Sec.~\ref{sec:structures_explicit}, this leads to the interpretation of the NHIM as the energy surface  of an invariant subsystem with one degree of freedom less than the full system (a supermolecule located for a frozen reaction coordinate (and conjugate momentum) between reactants and products).

Since, typically (for more than $1$ DoF), there is more than one
bath mode present, the total energy can be distributed in a variety of ways amongst the bath modes.  For example, putting all of the energy
into any one of the bath modes results in a single periodic orbit (often referred to as a \emph{Lyapunov orbit}).
More generally, the supermolecule of fixed energy $h$ has 
energy distributed between all of the $n-1$ bath modes.  Since the
integrals of motion are constant on trajectories, the amount of energy in
each particular bath mode remains fixed during the motion.  Such a
molecule therefore undergoes quasiperiodic motion; it oscillates
independently in the $n-1$ different bath modes.  The motion of a
typical trajectory in the NHIM is quasiperiodic on an
$(n-1)$-torus. Heuristically, one can think of the NHIM as the invariant surface made up of trajectories with all possible energy distributions between the bath modes (and no energy in the reactive mode).

%%%%%%%%%%%%%%%%%%%%%%%%%%%%%%%%%%

\subsection{The NHIM $\sphere^3$ ($n=3$ DoF)}
\label{sec:3dNHIM}

As noted above, the NHIM for $2$-DoF systems is simply a periodic orbit.
This is an immediate consequence of the fact that there is only a single bath (or centre) mode.  
In this section we consider the next most complex case in some detail,  the $3$-DoF situation ($2$ bath modes), for which the NHIM is diffeomorphic to $\sphere^3$. However, first we will describe a way in which $\sphere^3$ can be visualized in lower dimensions. 
This is helpful from the point of view of visual intuition. 
Our discussion in this section owes a great deal to the nice report of \cite{chisholm}.

\medskip
\noindent
\paragraph{Geometrical Construction of $\sphere^3$ in Three Dimensions.}
\medskip
 
To begin with, first note that in the usual manner, to draw a 1-sphere (i.e. a circle), we need two dimensons, to draw a 2-sphere we need three dimensions, and to draw a 3-sphere we need four dimensions. We will describe a method of drawing the 3-sphere which requires only three dimensions. This particular representation will enable
us to make a clear picture of the dynamics on the NHIM.
The method can be viewed as a generalization of the
presentation of a two-dimensional sphere, $\sphere^2$, in terms of two hemispheres (topological 2--balls, $B^2$) and an equator (a onedimensional sphere, $\sphere^1$) that we discussed in Sec.~\ref{sec:model-hw}. 
For $\sphere^3$, one can form a similar representation by using two open
$3$-balls, $B^3$, to represent ``Northern'' and ``Southern''
hemispheres, and identifying points on their boundaries, $\sphere^2$, to
make an ``equator'';

\begin{equation}
\sphere^3 \equiv B^3_{\text{North}} \cup
\sphere^2_{\text{Equator}} \cup B^3_{\text{South}}.
\end{equation}

\begin{figure}[htb!]
\begin{center}
\includegraphics[angle=270,width=4in]{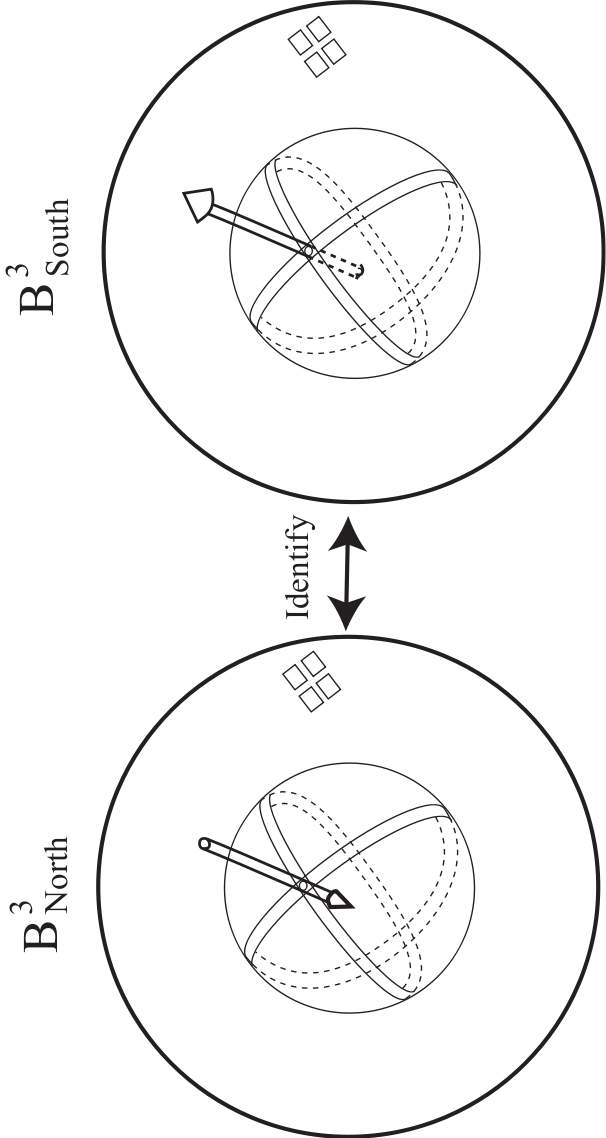}
 \caption{\label{fig:nhim-s3-model} A model of the sphere $\sphere^3$
formed by taking two open $3$-balls, $B^3_\text{North}$ and
$B^3_\text{South}$, and identifying points on their boundaries;
the ``equator'' $\sphere^2_\text{Equator}$. The arrow 
in $\sphere^3$ illustrates how one passes from the chosen ``North''
to the chosen ``South'' pole at the center of each $3$-ball.}
\end{center}
\end{figure}

\noindent
We will choose the  ``North'' pole (N) to be at the centre of the North ball and
the ``South'' (S) pole at the centre of the South ball.  
This model is illustrated in Fig.~\ref{fig:nhim-s3-model}.

\medskip
\noindent
\paragraph{Partitioning `Energy' Between the Two Bath Modes}
\medskip

Recall that (for $3$ DoF) there are $2$ bath modes.  For a given
trajectory in the NHIM,  all of the `energy' could be in the
first bath mode (resulting in a periodic orbit) or in the second bath
mode (resulting in another periodic orbit), or the
energy could be distributed, or partitioned, between the two bath modes.

\begin{figure}[htb!]
\begin{center}
\begin{tabular}{cc}
$B^3_\text{North}$& $B^3_\text{South}$\\
\includegraphics[width=1.4in]{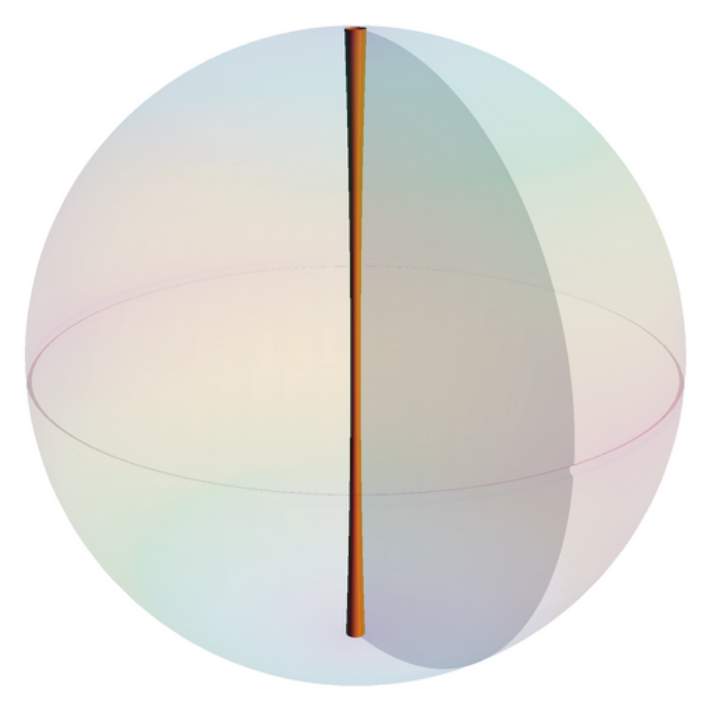}&
\includegraphics[width=1.4in]{torus000}\\
\end{tabular}
(a)\\
\begin{tabular}{cc}
\includegraphics[width=1.4in]{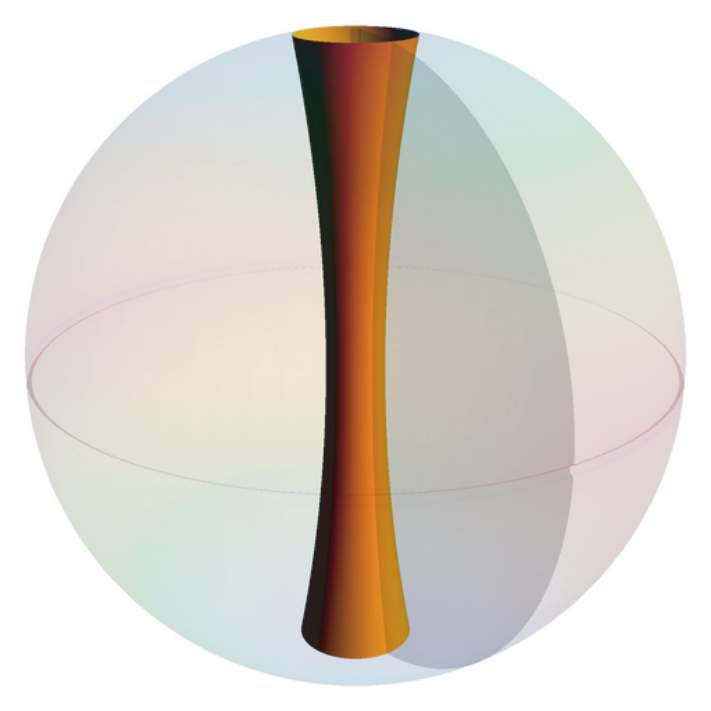}&
\includegraphics[width=1.4in]{torus010} \\
\end{tabular} 
(b) \\
\begin{tabular}{cc}
\includegraphics[width=1.4in]{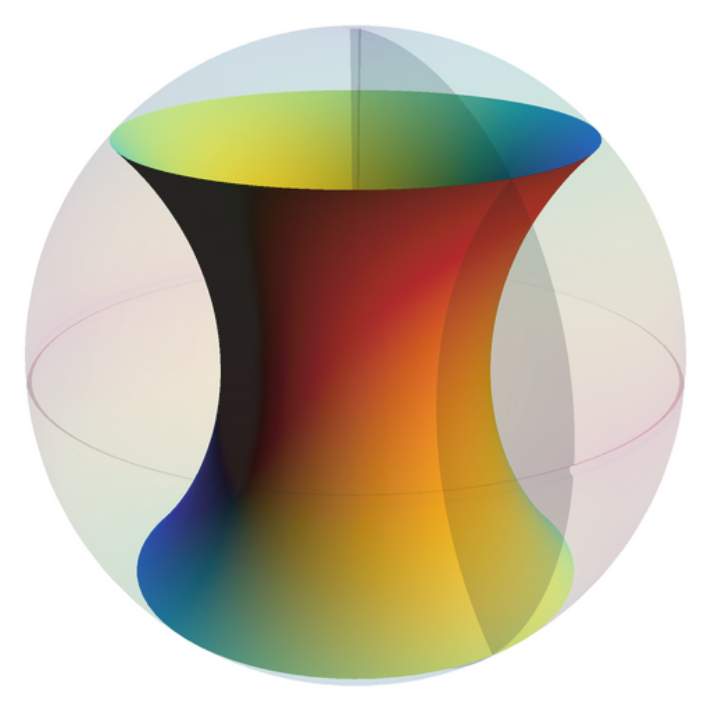}&
\includegraphics[width=1.4in]{torus050} \\
\end{tabular} 
(c) \\
\begin{tabular}{cc}
\includegraphics[width=1.4in]{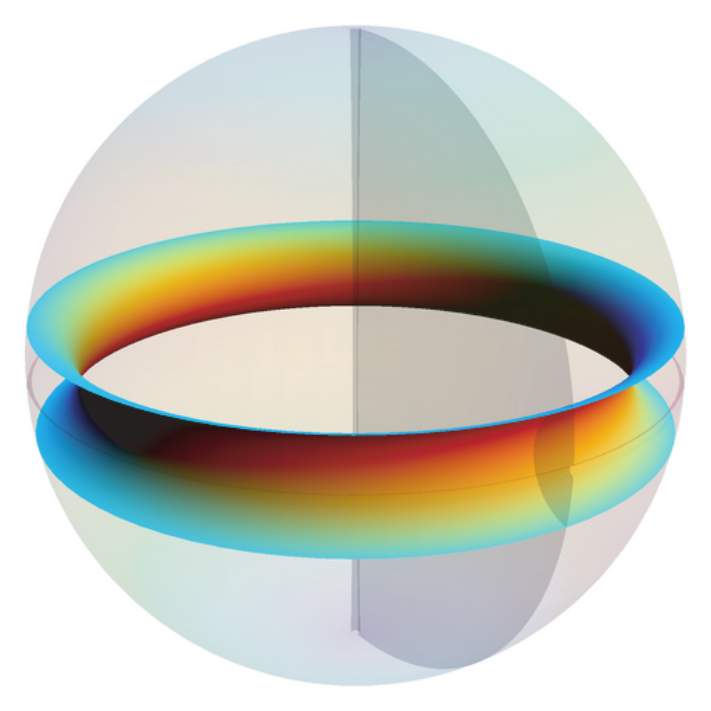}&
\includegraphics[width=1.4in]{torus090} \\
\end{tabular} 
(d) \\
\begin{tabular}{cc}
\includegraphics[width=1.4in]{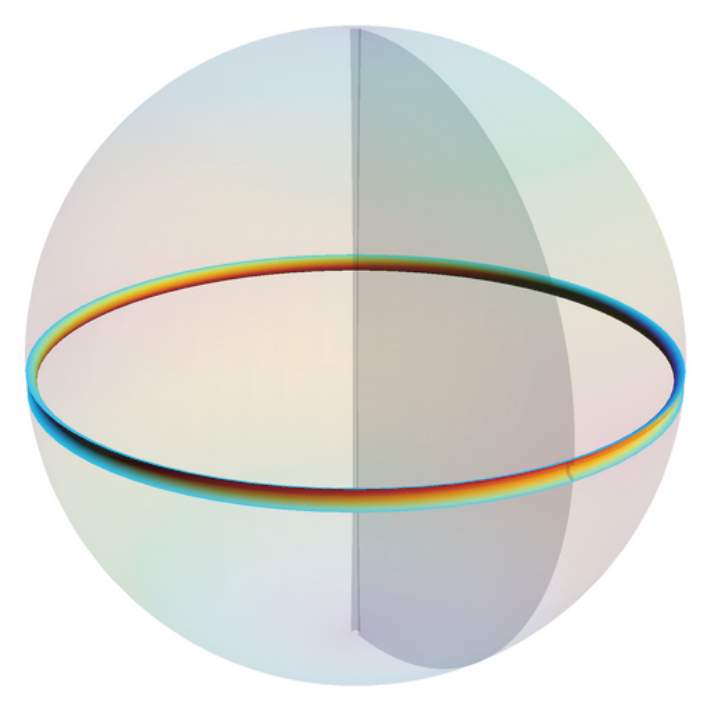}&
\includegraphics[width=1.4in]{torus099} \\
\end{tabular} 
(e) \\
\caption{\label{fig:nhim-tori}  Illustration of the invariant tori
in $\nhim$, for $n=3$ degrees of freedom. (a)
representation of the  periodic orbit corresponding to
local integrals of motion $(I,J_2,J_3)=(0,J_2^\text{max},0)$,
shown in both hemispheres of the model, (b) a nearby torus with
$J_3=\epsilon$  with
$0<\epsilon\ll J_2^\text{max}$, (c) the case $J_2=J_2^\text{max}/2$,
(d) a torus with $J_3=J_3^\text{max}-\epsilon$ which is close to
the periodic orbit corresponding to
$(I,J_2,J_3)=(0,0,J_3^\text{max})$, shown in (e).}
\end{center}
\end{figure}

In Fig.~\ref{fig:nhim-tori}(a) we show the two open balls that comprise our model of $\sphere^3$ as described above. We choose a trajectory having the property that all of the energy is in the first bath mode, i.e.,  $J_3=0$ and  $J_2=J_2^\text{max}$,  where $J_2^\text{max}$ is defined by the energy equation, $K_{\text{NF}}(0,J_2^\text{max}, 0)=h>0 $ (see Sec.~\ref{sec:structures_explicit}). This corresponds to a periodic orbit that moves along the ``vertical'' circle that passes through the centers of the two balls. As the periodic orbit exits the North ball at the top, it enters the South ball from the top, exiting the South ball at the bottom, and entering the North ball at the bottom.

Suppose now that we change this distribution of energy very
slightly, by taking a small amount of energy out of the first bath
mode and putting it into the second, which we illustrate in Fig.~\ref{fig:nhim-tori}(b) (and where we leave it to the reader to consider how a trajectory leaves one ball through its surface and enters the other ball). 
The motion still stays close to the original circle, but there is
now a second circular (oscillatory) motion ``around the periodic
orbit''. Thus the motion now takes place on a thin $2$-torus.  Any
initial condition on the $2$-torus corresponds to a quasiperiodic trajectory that remains on
the $2$-torus.

We continue this procedure of taking energy from the first bath mode and putting it into the second bath mode, until all of the energy is in the second bath mode (Fig.~\ref{fig:nhim-tori}(e)), which results in a periodic orbit moving in a ``horizontal'' circle along the surface of the two balls.

\begin{figure}[htb!]
\begin{center}
\includegraphics[width=2.9in]{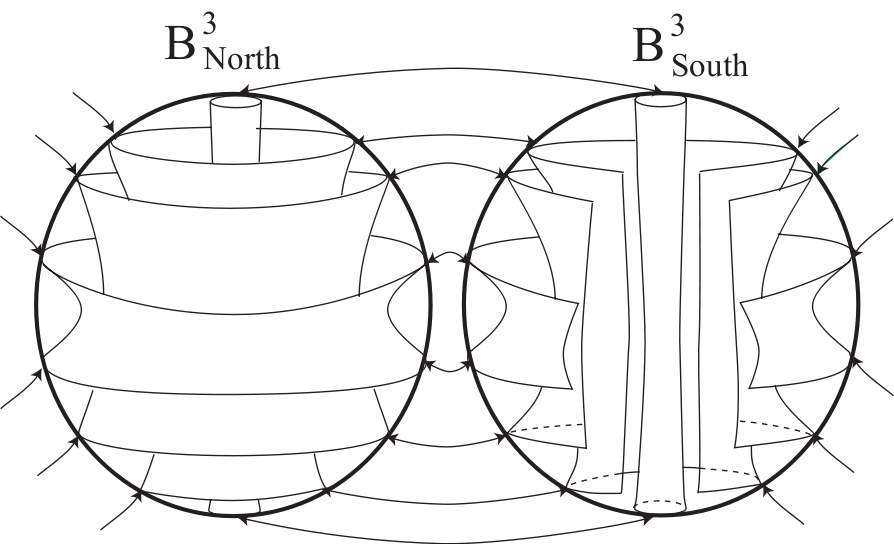}\\
\emph{$\leftarrow$ Identify $\rightarrow$}\\
\caption{\label{fig:nhim-tori-all}  The family of invariant $2$-tori that fill the NHIM, $\sphere^3_{\text{NHIM}}$, illustrated according to our method for visualizing $\sphere^3$ in three dimensions. One can think of each torus as being cut in half, with one half in the North ball and the other half in the South ball, and the ``edges'' of the two halves touching touching the surface of the $3$-ball, where they are identified. In this way, a quasiperiodic trajectory winding around a $2$-torus continually moves from one $3$-ball to another as it encounters the edge of the torus where the torus is cut into two pieces, which are on the surfaces of the $3$-balls (i.e. the ``equator'' of $\sphere^3_{\text{NHIM}}$). }
\end{center}
\end{figure}

In this way we see that $\sphere^3$ is foliated by a one-parameter family of 2-tori. Trajectories on the 2-tori correspond to quasiperiodic motion with energy contained in both bath modes. The ``extreme'' cases corresponding to all the energy contained in one of the bath modes corresponds to periodic orbits (the ``Lyapunov orbits''). We illustrate this family of 2-tori in Fig.~\ref{fig:nhim-tori-all}.

%%%%%%%%%%%%%%%%%%%%%%%%%%%%%%%%%%

\subsection{The NHIM, $\nhim$, for $n>3$ DoF}
\label{sec:ndnhim}

The method for visualizing the NHIM for 3 DoF described in the previous section is not as useful for 
$n>3$ DoF. Nevertheless, the dynamics on the NHIM have the same characteristics--quasiperiodic trajectories with the energy partitioned amongst the $n-1$ bath modes.  Hence, as mentioned in Sec.~\ref{sec:ndof}, for the general DoF case, a description of the NHIM can be given in terms of the geoemtry of the space of centre actions (or ``bath mode'' actions) $J_2,\ldots,J_n$. This is what we will now describe.

Classically, the space formed by the centre actions is the
$(n-1)$-dimensional \emph{positive cone}; equivalently, the
positive orthant  (the generalization  of a quadrant in the plane or an octant in three dimensions to arbitrary dimensions) 
given by $J_2,\ldots,J_n\ge 0$ in $\mathbb{R}^{n-1}$.  In
general, the NHIM is represented in this space by a (topological)
$(n-2)$-simplex having $(n-1)$ vertices, one for each centre
action.  The vertices are always located on the coordinate axes;
each vertex thus corresponds to $(n-2)$ of the integrals $J_k$
being equal to zero.  Note that the projection of the NHIM to this
space of actions will generally be a \emph{nonlinear} embedding of
a simplex; the embedding is only linear (that is, all cells of the
simplex are \emph{flat}) in the case where the vector field
defined by Hamilton's equations is linear.  We will denote this
embedded simplex by

\begin{equation}
\Lambda=[2,3,\ldots,n],
\end{equation}

\noindent
using the index of each centre action to label a vertex.  Within
this simplex, each $(k-1)$-cell (for every $k=1,\ldots,(n-1)$),
denoted $[j_1,\ldots,j_{k}]$, represents a $(k-1)$-parameter
family of invariant $k$-tori which we denote by

\begin{equation}
\mathbb{T}^{k}_{j_1j_2\ldots j_{k}}.
\end{equation}
Here $j_1,,\ldots,j_{k}$ are mutually different elements of $\{2,\ldots,n\}$ which indicate the center actions $J_{j_1},\ldots,J_{j_k}$ are different from zero. 
The family is parameterised, for example, by the centre actions
$J_{j_1},\ldots,J_{j_{k-1}}$, with the value of $J_{j_{k}}$ being
determined by the energy equation, as we will describe
below.

For example, each vertex (i.e., $0$-cell), denoted $[j]$ (for
$j=2,\ldots,n$), represents a single periodic orbit
$\mathbb{T}^1_{j}\cong \sphere^1$, in which all actions except $J_j$ are set to
zero and, therefore, the remaining action is fixed at a value
$J_j=J_j^\text{max}$, which solves the following equation that
corresponds to restriction to a single energy surface,

\begin{equation}
K_\text{NF}(I,J_2,\ldots,J_n)=h>0,
\end{equation}

\noindent
subject to setting ${J_{\ell}}=0$ for all $\ell=2,\ldots,n$, $\ell\neq j$.  The
periodic orbit, $\mathbb{T}^1_j$, 
projects to a single circle of radius $\sqrt{J_k^\text{max}}$ in
the $(q_k,p_k)$-plane of the normal form coordinates and to the
origin in all other normal form planes, $(q_\ell,p_\ell)$ for
$\ell\neq j$.

Each edge $[j_1,j_2]$ (for $j_1\neq j_2$) of the simplex
represents a $1$-parameter family of invariant $2$-tori, in which
all actions except $J_{j_1}$ and $J_{j_2}$ have been set to zero.
Without loss of generality, we can choose $J_{j_1}$ as a
parameter; each value $J_{j_1}\in[0,J_{j_1}^\text{max}]$ fixes the
value of $J_{j_2}$ by virtue of the above iso-energetic condition
subject to the modified restriction that

\begin{equation}
J_{\ell}=0\ \mbox{for $\ell\neq j_1,j_2$}.
\end{equation}

\noindent
We denote this $1$-parameter family of invariant $2$-tori in
$\nhim$ by $\mathbb{T}^2_{j_1j_2}$.

By allowing $k$ centre actions to be non-zero in the NHIM,
corresponding to a $(k-1)$-cell $[j_1,j_2\ldots,j_{k}]$ of the
simplex, we can, without loss of generality, take
$J_{j_1},J_{j_2},\ldots,J_{j_{k-1}}$ as parameters which fix the
value of $J_{j_{k}}$ via the condition

\begin{equation}
H_\text{NF}(I,J_2,\ldots,J_n)=h>0,
\end{equation}

\noindent
subject to setting

\begin{equation}
{J_{\ell}=0\ \mbox{for $\ell\neq j_1,j_2,\ldots,j_{k}$.}}
\end{equation}

\noindent
This defines a $(k-1)$-parameter family of invariant $k$-tori,
denoted $\mathbb{T}^{k}_{j_1j_2\ldots j_{k}}$.

\begin{figure}[htb!]
\begin{center}
\includegraphics[width=1.75in]{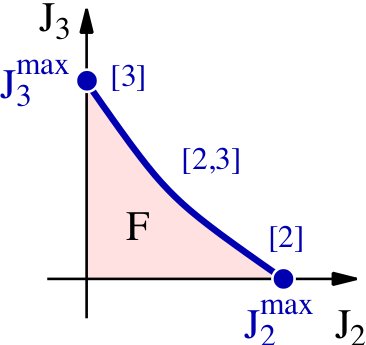}\\
(i) $n=3$ degrees of freedom.\\
\includegraphics[width=2.0in]{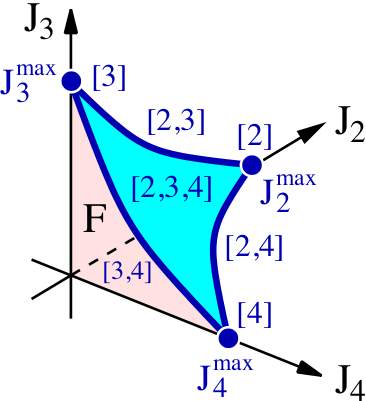}\\
(ii) $n=4$ degrees of freedom.\\
\includegraphics[width=2.75in]{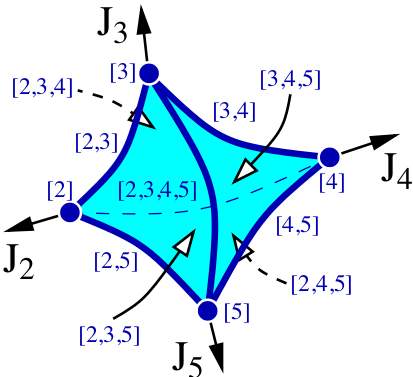}\\
(iii) $n=5$ degrees of freedom. 
\caption{\label{fig:nhim-actions} Schematic representation
of the projection of $\nhim$ into the space of centre
actions, $J_2,J_3,\ldots,J_n$, for $n$  DoF giving a nonlinear embedding of
a simplex with $(n-1)$ vertices, $[j]$, $j=2,\ldots,n$, which encloses a volume proportional to the directional flux, $F$.}
\end{center}
\end{figure}

A schematic representation of the projection of $\nhim$
into the space of centre actions $J_2,J_3,\ldots,J_n$ for $n$-DoF
is shown in Fig.~\ref{fig:nhim-actions}  for (i) $n=3$, (ii) $n=4$, (iii) $n=5$.   
In (iii) we show a schematic $3$-dimensional projection
of the $4$-dimensional space of centre actions.

\medskip
\noindent
\paragraph{Directional flux across the dividing surface.}
\medskip

In \cite{WaalkensWiggins04} it is shown that the (directional) flux
across a given iso-energetic dividing surface $\ts$ (that
is, the equal and opposite fluxes through the forward and backward
dividing surface hemispheres $\tsf$ and $\tsb$) may be reduced to a certain integral
taken over $\nhim$.  In the representation of the NHIM in
in terms of the centre integrals given above, this flux corresponds to the
$(n-1)$-dimensional volume of the finite region of the positive
orthant bounded by the $(n-2)$-simplex representing the NHIM times $(2\pi)^{n-1}$.
That is up to a factor, the flux corresponds to the volume of an $(n-1)$-simplex
formed by adding the origin to the vertex set of the NHIM in the
space of integrals. In Fig.~\ref{fig:nhim-actions}  for 
(i) and (ii), the $(n-1)$-dimensional volume $F$ corresponding to the
magnitude of the directional flux across the dividing surface
hemispheres $\tsf$ and $\tsb$ is
indicated. Applications of this in both the classical and quantum case can be found in 
\cite{WaalkensSchubertWiggins08}.

% end rem

%%%%%%%%%%%
%%%%%%%%%%%
%\input{conclusions}
%%%%%%%%%%%
%%%%%%%%%%%

%---------------------------------------------
\section{Conclusions}
\label{sec:conc}
%----------------------------------------------

This paper is concerned with strategies and approaches for visualizing the phase space structures that govern reaction dynamics. In 
Section~\ref{sec:summary_phases_space_structures} we reviewed recent results that develop the phase space geometry near an equilibrium point of Hamilton's equations of $\mbox{saddle} \times \mbox{centre} \times\cdots\times
\mbox{centre}$  stability type (`saddle' for short) in $n$ degree-of-freedom Hamiltonian systems.  The key technique used was the method of Poincar\'e-Birkhoff normal forms. This method provided a transformation (and its inverse), valid in a phase space neighborhood of the equilibrium point,  to normal form coordinates in which the phase space structures could be explicitly realized (and then mapped back to the original coordinates via the inverse of the normal form transformation). The key phase space structure was seen to be a normally hyperbolic invariant manifold (NHIM), which is the higher dimensional analogue of a ``saddle point''. The NHIM is the anchor upon which a dividing surface having the ``no-recrossing'' property and minimal flux is constructed. Moreover, the stable and unstable manifolds of the NHIM are the  ``conduits'' 
for forward and backward reactive trajectories (trajectories evolving from reactants to products, and vice-versa). More precisely, one the two branches of the stable manifold  can be 
glued together with one of the two branches of  the unstable manifold to form a  cylinder (denoted the \emph{forward reactive cylinder} in this paper) that enclose all forward reactive trajectories; and the other branches of the stable and unstable manifolds can be glued together to form a cylinder (the \emph{backward reactive cylinder}) which encloses all backward reactive trajectories.    
In the case where a generic non-resonance condition holds on the  pure imaginary eigenvalues associated with the center directions it follows that in the normal form coordinates the Hamiltonian  (when the series  representing the normal form Hamiltonian is truncated at some finite order) has $n$ integrals of the motion and thus is integrable. This integrable structure allows for a natural decoupling of  the dynamics into a single reactive mode and $n-1$ bath modes. Moreover, the  integrable structure provides a foliation of the NHIM  by a family of lower dimensional tori and a foliation of a neighborhood of the equilibrium point into Lagrangian cylinders. 

The ``decoupling'' of the  vector field and the associated dynamics masks the complete picture of reaction governed by the phase space structures in the energy surface. In Section~\ref{sec:2dofModels} we provide three different 2 degree-of-freedom models that allow us to visualize the energy surface, the NHIM and its stable and unstable manifolds, the dividing surface, and the Lagrangian cylinders.  
These give a complete picture of how the phase space structures  govern the dynamics of trajectories near a saddle.
By contrast,  the role of the phase space structures is typically obscured in their  common representation as projections to configuration. For example, the projection of the volume enclosed by the forward and backward cylinders might not lie inside of the projection of the forward and backward cylinders, and as a consequence forward and backward reactive trajectories might seemingly  leave the forward and backward reactive cylinders 
(for an example, see the study of the planar Hill problem in \cite{WaalkensBurbanksWiggins05b}).  

For the general case of $n>2$ degrees of freedom systems,  we use the integrals alone to provide a  model for visualizing the  phase space structures (see Section~\ref{sec:ndof} ). This model does not contain explicit information on trajectories, but it can be easily related to the behavior of trajectories by utilizing the foliation of the neighborhood of the NHIM by  Lagrangian cylinders. 

We moreover discussed in some detail the foliation of the  NHIM for 3 degrees of freedom.  Here the NHIM is a threedimensional sphere, $\sphere^3$ that is foliated by a one-parameter family of invariant 2-tori. We visualized this foliation using the 
3-ball model of $\sphere^3$ (see Section~\ref{sec:nhim}). For the general case of $n>3$ degrees of freedom, we  presented the NHIM as a simplex in the space of the integrals. This is of particular importance, since, up to a prefactor, the volume enclosed by the simplex in the positive orthant of the integral space gives the directional flux.  

We conclude by emphasizing that we focused the discussion of the phase space structures and their visualization to the neighborhood of the saddle.
In fact, depending on the global topology of the energy surface,  the local partitioning of the energy surface by the dividing surface constructed from the NHIM into a reactants and a products component might not extend to a global partitioning. For example, there could be more than one saddle point which controls the access to and escape from a certain phase space region which requires a more careful definition of a reactants and a products region.  Moreover, in this paper we did not discuss the global geometry of the stable and unstable manifolds of the NHIM. In fact the stable and unstable manifolds might extend to regions far away from the neighborhood of the saddle.  Even if there is only a single saddle point, the global structure might be very complicated since the stable and unstable manifolds might wrap around in a complicated manner and intersect each other to form  a highdimensional  homoclinic tangle. If there are several saddles the stable and unstable manifolds of NHIMs of different saddles might intersect.  Since the stable and unstable manifolds are of codimension 1 in the energy surface they still act as impenetrable barriers globally, and also control the global dynamics in a possibly complicated way. The details of the dynamics then strongly depends on the system under consideration. For a few examples, we refer to \cite{WaalkensBurbanksWiggins04,WaalkensBurbanksWigginsb04,WaalkensBurbanksWiggins05,WaalkensBurbanksWiggins05b,WaalkensBurbanksWiggins05c}.

% end rem

\section{Acknowledgements}

SW  acknowledges the support of the  Office of Naval Research Grant No.~N00014-01-1-0769,
and the stimulating environment of the NSF sponsored Institute for
Mathematics and its Applications (IMA) at the University of Minnesota,
where this manuscript was completed.
HW acknowledges support by EPSRC under Grant No. EP/E024629/1.
We are grateful to Dr. Andrew Burbanks for helping with an early version of this manuscript. 
%and for producing some visualisations using the Visualisation Toolkit
%(VTK) , the POVRay Raytracer, and xfig. 
%We thank Jonathan Robbins for a careful reading of an early  draft
%manuscript.

%\bibliography{chembib_geom_paper}

\def\cprime{$'$}

\end{document}